\documentclass[twocolumn]{aastex631}

\accepted{December 24, 2023}

\submitjournal{ApJ}

\graphicspath{{./}{}}
\begin{document}

\title{Dust around massive stars is agnostic to galactic environment: New insights from PHAT/BEAST}

\correspondingauthor{Christina Willecke Lindberg}
\email{clindbe2@jhu.edu}

\author[0000-0003-0588-7360]{Christina Willecke Lindberg}
\affiliation{Johns Hopkins University, 3400 North Charles St., 473 Bloomberg Center for Physics and Astronomy, Baltimore, MD, 21218}

\author[0000-0002-7743-8129]{Claire E. Murray}
\affiliation{Space Telescope Science Institute, 3700 San Martin Drive, Baltimore, MD, 21218}
\affiliation{Johns Hopkins University, 3400 North Charles St., 473 Bloomberg Center for Physics and Astronomy, Baltimore, MD, 21218}

\author[0000-0002-1264-2066]{Julianne J.\ Dalcanton}
\affiliation{Center for Computational Astrophysics, Flatiron Institute, 162 Fifth Ave, New York, NY 10010, USA}
\affiliation{Department of Astronomy, Box 351580, University of
  Washington, Seattle, WA 98195}

\author[0000-0003-4797-7030]{J. E. G. Peek}
\affiliation{Space Telescope Science Institute, 3700 San Martin Drive, Baltimore, MD, 21218}
\affiliation{Johns Hopkins University, 3400 North Charles St., 473 Bloomberg Center for Physics and Astronomy, Baltimore, MD, 21218}

\author[0000-0001-5340-6774]{Karl D. Gordon}
\affiliation{Space Telescope Science Institute, 3700 San Martin Drive, Baltimore, MD, 21218}

\begin{abstract}

% Literature Review/Background Information
% Massive stars are bright, short-lived stars which, due to their brief lifespans, are found to broadly trace the interstellar medium. Massive stars should be capable of forming anywhere with sufficient densities of molecular gas, however the length-scale of clouds capable of forming massive stars is still unconstrained. 

 %Massive stars trace the interstellar medium (ISM) from which they were born. Although we expect most massive stars to remain near their birth sites, stellar feedback, turbulent evolution in the ISM, and natal kicks can rapidly change the stars' immediate environment.
% Research Gap
%Extragalactic observations of direct molecular gas tracers are often tens of parsecs in spatial resolution, which limits their ability to probe to small-scale structures of the ISM. 

Resolving the environments of massive stars is crucial for understanding their formation mechanisms and their impact on galaxy evolution. An important open question is whether massive stars found in diffuse regions outside spiral arms formed in-situ or migrated there after forming in denser environments. To address this question, we use multi-resolution measurements of extinction in the Andromeda Galaxy (M31) to probe the ISM surrounding massive stars across galactic environments.

We construct a catalog of 42,107 main-sequence massive star candidates ($M \geq 8 M_{\odot}$) using resolved stellar photometry from the Panchromatic Hubble Andromeda Treasury (PHAT) program, plus stellar and dust model fits from the Bayesian Extinction and Stellar Tool (BEAST). We quantify galactic environments by computing surrounding stellar densities of massive stars using Kernel Density Estimation. We then compare high-resolution line-of-sight extinction estimates from the BEAST with 25-pc resolution dust maps from PHAT, measuring the total column density distribution of extinction.

Our key finding is that, although the average total column density of dust increases with the density of massive stars, the average line-of-sight extinction towards massive stars remains constant across all environments. This suggests that massive stars have a uniform amount of dust in their immediate environment, regardless of their location in the galaxy. One possible explanation for these findings is that small molecular clouds are still capable of forming massive stars, even if they are not resolvable at 25-pc. These results indicate that massive stars are forming in the sparse regions of M31, as opposed to migrating there.

\end{abstract}

\keywords{Stellar populations --- Massive stars --- Interstellar medium --- Interstellar dust ---Hubble Space Telescope}

% common paper citation aliases
\defcitealias{williams2014}{W14}
\defcitealias{gordon2016}{G16}
\defcitealias{dalcanton2015}{D15}

\section{Introduction} \label{sec:intro}

% where do massive stars form?
Massive stars are generally thought to form within giant molecular clouds (GMCs) and molecular cloud complexes, where enough star-forming material can coalesce to fully sample the initial mass function (IMF) \citep{ladalada2003}. While observations broadly corroborate this theory, there is still considerable debate regarding which formation mechanism dominates the formation of massive stars, e.g. competitive accretion \citep{Zinnecker1982} versus core/monolith collapse \citep{shu1987}, and thereby what environments are conducive to the formation of massive stars i.e. could massive stars form in relative isolation in less gas-rich environments in galaxies \citep{lamb2010}? 

% While massive stars have been detected outside dense star-forming regions in galaxies, it is often the case that these sources are runaway stars, exhibiting large proper motion velocities relative to other young stars. Massive stars are particularly susceptible to having runaway kinematics due to high rates of forming in clusters or binary systems \citep{blaauw1961, lamb2016}, which can result in them being catapulted/kicked away from their natal environment. As a result, there is limited evidence of massive star formation outside of star-forming regions. Some of the only evidence so far comes from recent observations in the Small Magellanic Cloud, which found 14 massive stars formed in relatively isolated locations \citep{oey2013}. 

% Understanding how massive stars influence their local interstellar medium across a variety of galactic environments is critical for properly modeling the evolution of molecular clouds and predicting subsequent star formation in galaxies. 

% why do we care about where they form?
Knowing where massive stars can form is critical for understanding the evolution of the interstellar medium (ISM) and galaxies. Massive stars are short-lived and highly energetic, emitting feedback into the surrounding ISM. Supernovae have traditionally been viewed as the most influential and destructive feedback mechanism. However, massive stars also emit multiple types of pre-supernovae feedback such as stellar winds, ionizing ultra-violet (UV) radiation, UV radiation pressure, reprocessed infrared (IR) radiation, and cosmic rays, in addition to supernovae, all of which can cumulatively influence or destroy the ISM over several million years (Myrs) \citep[for more details, see chapter by][]{Chevance2023}.

Energy from stellar feedback mechanisms has the potential to either (a) heat up and inject turbulence into the surrounding ISM, suppressing subsequent star formation \citep{maclow2004}, or (b) sweep up star-forming material into colliding flows, increasing the local star formation rate \citep{dobbs2020}. The first effect is evidenced by the fact that a significant fraction of young stellar clusters have no molecular gas associated with them \citep{kruijssen2019}, the result of pre-supernova stellar feedback from massive stars destroying the surrounding GMCs within short timescales (3-5 Myrs) \citep{rosen2022, Chevance2022}. On the other hand, 3D dust mapping and observations of young stellar objects in the Local Bubble have shown that energy from multiple supernovae can create a shell of molecular gas, fueling subsequent star formation \citep{zucker2022}. At the moment, it is unclear which of these two opposing effects dominates star-formation rates on a galactic scale, however, studies using simulations seem to show that the effects of stellar feedback might be dependent on the density of the local ISM around stars \citep{dobbs2022, gatto2015}.

While simulations have given us the mechanism to understand how various forms of feedback interact with the ISM on small scales \citep{dale2014, peters2011, geen2015, kim2017, chen2015, choi2017}, it is still difficult to model the effects of feedback across entire galaxies \citep{girichidis2016, dobbs2022}. Modeling feedback in detail requires sub-parsec (pc) scale resolution \citep[e.g. TIGRESS, ][]{changgoo2017}, making it difficult to propagate the effects to the kpc-scales of galaxies \citep[e.g. FIRE, ][]{hopkins2014}. 

% To better understand how stellar feedback influences star formation across galactic environments, we first need a better picture of what the local ISM looks like around massive stars. Feedback has different effects depending on how deeply embedded a star is in the ISM.

% More specifically, cold molecular gas which is the predecessor to star formation. Despite being the most abundant form of molecular gas, molecular hydrogen ($H_2$) can usually only be measured at temperatures higher than will star formation (T$\sim 100$ K) due to its high dipole excitation temperature. Therefore, to approximate the total amount of \textit{cold} molecular gas in a region, we rely on interstellar dust grains to help us measure the local ISM. Interstellar dust grains are generally well-mixed with interstellar gas and therefore trace a large range of gas densities, from atomic hydrogen to dense molecular clouds. This is advantageous compared to other molecular gas traces, such as CO and CN, which often have a lower ionization temperature than $H_2$, which can result in underestimates of the total molecular gas mass due to the presence of "CO-dark" gas (cite). In addition, due to current telescope limitations, molecular gas maps of other galaxies are often limited to resolutions $\sim10$-pc, too large to effectively probe pre-SNe feedback effects (cite). 

In this paper, we use interstellar dust to directly probe the local ISM around stars. Interstellar dust is well-mixed with gas and traces a wide range of gas densities. Due to its extinctive properties, we can use multi-band photometry of resolved stars to measure the column density of dust directly along the line of sight. This gives us the ability to not only probe the local ISM around individual stars, but to also use aggregate extinction statistics from older stellar populations to measure the total column density of dust in nearby galaxies. These two measurements give us unique information about how stars are situated within the ISM, allowing us to characterize the embeddedness of massive stars.

% Stars form within molecular clouds (GMCs) and molecular cloud complexes. We may naively expect the stars to remain embedded in these natal clouds, if the only relevant evolution was driven by the stars' velocity dispersion, which has been measured to be on the order of a couple km s$^{-1}$ \citep{swiggum2021, galli2019}.  Empirically, we see ample evidence that these effects may be in play, especially in active star-forming regions such as...(observations A, B, C). But we don't know how these change across galactic environments...

% Transition sentences here. New paragraph about how we can gain some unique information by looking at dust which traces the cold molecular phase. Point out that dust gives you unique information about how a star is situated within the ISM, that we can't get from just a map of CO/CN, etc. 

% If dispersion were the only factor influencing how embedded stars are in the ISM, we would expect massive stars to remain relatively embedded given their short lifespans. However,

To ensure we are sampling from a wide range of galactic environments, we construct a photometric catalog of main-sequence massive star candidates in the Andromeda Galaxy (M31) and compare how the stellar line-of-sight extinction and average total column density extinction differ across regions of M31. We derive high-resolution ($\sim$0.5 pc) line-of-sight extinction measurements from \citet[hereafter \citetalias{gordon2016}]{gordon2016} which probe the immediate surroundings of each star.
For comparison, we use 25-pc-resolution extinction maps from \citet[hereafter \citetalias{dalcanton2015}]{dalcanton2015} to characterize the total regional column density of dust. In conjunction, these two measurements provide context for how the amount of dust varies as a function of resolution at scales typically not probed by other gas tracers in extragalactic environments.

% What is the structure of this paper? 
In Section \ref{sec:data}, we introduce our extinction data sets and discuss the selection of our massive star candidates. We remove potential contaminants in Section \ref{sec:catalog} and compare our candidate catalog with other massive star catalogs in Section \ref{sec:valid}. In Section \ref{sec:results}, we characterize the underlying stellar density of massive stars and show how the two extinction resolutions vary as a function of stellar density. In Section \ref{sec:disc}, we discuss the implications of our results and compare our findings to other tracers such as CO and HI. In Section \ref{sec:followup}, we discuss potential follow-up observations for confirming our results, and, in Section \ref{sec:conc}, we summarize our findings

\section{Data} \label{sec:data}

% Give a brief overview of the PHAT program and how the data was acquired (HST, WFC3 photometry, UV to IR). 
The Panchromatic Hubble Andromeda Treasury (PHAT) is a multi-wavelength survey of the disk of M31, providing an unprecedented opportunity to study the structure and formation history of M31 \citep{dalcanton2012, williams2014}. PHAT observed roughly 1/3 of the star-forming disk of M31 from July 2010 to October 2013 using the Hubble Space Telescope (HST) Advanced Camera for Surveys (ACS) Wide Field Channel (WFC), Wide Field Camera 3 (WFC3) ultra-violet and visible (UVIS) channels, and WFC3 infrared (IR) channel. PHAT data products and photometry can be retrieved from the Mikulski Archive for Space Telescopes \citep[MAST,][]{PHATdoi}. The survey was designed to efficiently image a large portion of M31, with a focus on the northeast quadrant due to its low apparent dust extinction and limited contamination from M32, a nearby dwarf satellite. The survey area is composed of 23 ``bricks", where each brick is composed of 18 fields corresponding to the WFC3 IR camera footprint. Each field was observed in six bands, spanning the near-UV (F275W, F336W), visible (F475W, F814W), and near-IR (F110W, F160W). These filters were selected to minimize wavelength overlap and maximize the variety of observable stars: near-IR for stars in dusty regions and near-UV for hot stars.

To identify individual sources, they used DOLPHOT to measure point-spread function (PSF) photometry from stacks of PHAT images \citep{dolphin2002}. Parameters such as sharpness (i.e., how much the flux of the central pixel deviates from the expected PSF) and crowding (i.e., how many magnitudes of flux were removed from overlapping stars), were used to eliminate contaminants such as cosmic rays, background galaxies, and stellar blends \citep[see][Table 3]{williams2014}. WFC3 UVIS and IR channels have resolutions of 0.04$^{\prime \prime}$/pix and 0.13$^{\prime \prime}$/pix, respectively. Over 100 million stars were resolved using UVIS F814W. Completeness and uncertainty were assessed using 5.4 million artificial star tests (ASTs) per brick. For outer regions of the disk, the photometric depth was determined to be F475W $\sim$ 28 mag, defined by the 50\% completeness limit, the magnitude at which 50\% of the ASTs are no longer recovered. However, the photometric depth is much shallower in the inner regions of the disk near the bulge, F475W $\sim$ 25 mag, due to crowding in the visible and NIR.  

The inferred stellar parameters described in this work are based on the \citet{williams2014} photometry. Recently, the photometry pipeline for PHAT was reworked to use all overlapping data and updated charge transfer efficiency (CTE) correction to allow for better detection of faint sources \citep{williams2023}. This resulted in a systematic increase of $\sim$0.05 mag in the NUV bands, however, there were no other significant changes to the photometry for sources brighter than 25 mag. Since all massive star candidates identified in the final catalog are brighter than 22 mag in NUV filters (Figure \ref{fig:CMD_final}), we do not expect a significant change in their photometry. %It is possible that improved NUV photometry could result in more detections of highly embedded massive stars, however, it is unclear whether this would have any significant impact on the final result. Regardless, rerunning the BEAST SED fitting procedure to characterize the line-of-sight extinction and stellar initial mass for the new PHAT legacy photometry would be incredibly computationally expensive, warranting its own publication, and is therefore outside the scope of this paper.

\subsection{Average Regional Extinction}\label{sec:Av25} 

We use high-resolution 25-pc dust extinction maps of M31 to characterize the average dust content in the neighborhood around each massive star, $A_{V,\, 25}$. Using groups of $\sim$100 red giant branch (RGB) stars, \citetalias{dalcanton2015} modeled the NIR color-magnitude diagram (CMD) as a combination of an unreddened foreground and a background population of stars observed through a complex layer of dusty gas. The reddened RGB stars are assumed to sample a log-normal distribution of dust columns, confined to a thin midplane embedded in a much thicker distribution of RGB stars \citep[see][]{dalcanton2023}. As such, the resulting extinction maps characterize the statistical properties of the integrated column density of dust but do not inform the exact extinction of any individual star. The massive stars in this study, for example, generally span a scale height similar to molecular gas in the ISM ($\sim$ 100 pc), meaning they are more likely to be intermixed with the dusty ISM \citep{patra2019, meng2021}. %As a result, they may not actually sample the full column density of dust. %They may also sample higher or lower extinction lines of sight within the 25~pc analysis region.

There are two caveats to keep in mind for the use of the \citet{dalcanton2015} maps in this paper. First, this method does not accurately characterize low levels of extinction; for $A_V < 0.5$, reddening is too small to broaden the RGB at a level that can be accurately measured using a CMD. As such, we chose to omit any stars from our catalog if they occupy regions of $A_{V, 25} < 0.5$. The influence of this omission is discussed in Section \ref{sec:variable_av}. Second, because the maps are based on assuming a log-normal distribution of dust columns within each pixel, they are not sensitive to the additional power-law tails to high column densities that have been seen in some star-forming molecular clouds \citep[e.g.,][]{Kainulainen2014}. These very high-density cores cover a tiny fraction of the area and thus are easily missed by the limited number of background RGB stars sampling the dust distribution. However, such regions may be preferentially sampled by newly formed massive stars, and thus we would not be surprised if some massive stars had extinctions that statistically favored unusually high extinctions.
 
\subsection{Line-of-Sight Extinction}

% theory introduction
We derive line-of-sight extinction measurements for each star ($A_{V,\,LOS}$) using their spectral energy distributions (SED), as observed by the six-filter photometry from PHAT. The observed SED of a star is the result of four main processes: (1) a star emits photons in the form of blackbody radiation over a large range of wavelengths, and the stellar photosphere imprints a complex signature on the spectrum via absorption lines; (2) intervening dust between Earth and the star scatters and absorbs a fraction of those photons; (3) the transmitted photons with frequencies inside the filter wavelength range impact the detectors of our telescopes; and (4) we use point spread functions to distinguish photometry from different stars, a process which is limited and biased by the presence of blended neighboring stars, both detected and undetected. %\footnote{The influence of the third process, i.e. crowding, can be quantified using artificial star tests (ASTs), where simulated stars are injected into the raw data and reprocessed by the photometry pipeline. If stars below a certain magnitude are only reobserved/recovered 50\% of the time, this magnitude is then deemed the magnitude limit at which our catalog can be considered complete.}. 
Dust preferentially extinguishes bluer/shorter wavelengths, meaning that if we obtain UV-IR photometry, it is possible to break the degeneracy between the observational effects of stellar temperature and extinction, especially when all stars can be assumed to be at a common distance. 

% The initial mass of a star can be derived from its SED, assuming its flux is unextinguished, and its distance is well-constrained. However, almost all stars have some dust along their line-of-sight, meaning that any stellar fits must also account for extinction. Extinction estimates are hard to quantify because stellar temperature and dust both decrease the observed flux of a star. However, due to the distribution of grain sizes, 

Using the wide wavelength coverage from PHAT (near-UV to near-IR), \citetalias{gordon2016} obtained stellar models and line-of-sight extinction models for over 40 million sources in M31. These sources came from the ``good star" (GST) catalogs, meaning they had to be detected in at least four bands and pass particular quality standards based on their PSF photometry \citep[for details, see Table 3 in][]{williams2014}. Model fits were obtained using the Bayesian Extinction and Stellar Tool (BEAST)\footnote{https://github.com/BEAST-Fitting/beast}, a probabilistic tool for modeling sources in resolved stellar surveys. The BEAST is an open-source \texttt{PYTHON} package designed to infer the intrinsic properties of resolved sources using stellar evolution and atmosphere models in combination with line-of-sight dust extinction models. Using a Bayesian probabilistic grid search, the BEAST parameterizes the SEDs of individual stars by simulating a variety of stellar and extinction models. First, stellar models are sampled from stellar evolutionary tracks, generating a range of stellar parameters: initial mass ($M_{ini}$), age ($logA$), and metallicity ($Z$). Stellar models are generated using stellar evolutionary tracks from Padova \citep{marigo2008, girardi2010} or PARSEC \citep[][]{bressan2012}. Then, stellar atmosphere grids are used to simulate spectra. The BEAST uses two stellar atmosphere grids, the local thermal equilibrium (LTE) CK04 grid \citep[][]{castelli2003} and the non-LTE TLusty OSTAR and BSTAR grids \citep{lanz2003, lanz2007} to cover a large range of stellar types. Afterward, variable amounts of extinction are applied to the spectra, determined by the range of extinction parameters: dust column densities ($A_V$), total to selective extinction ($R_V$), and a mixture model between the Milky Way and SMC-like extinction curve shapes ($f_A$). Finally, to convert the model extinguished spectra to model extinguished SEDs, the extinguished spectra are integrated over the observed filter bandpass functions to account for the impact of dust extinction in broad bandpasses. This produces a large grid of extinguished stellar SEDs covering the full range of pre-determined stellar ($M_{ini}$, $log\,A$, $Z$) and dust parameters ($A_V$, $R_V$, $f_A$). From these parameters, we derive secondary stellar parameters such as effective temperature ($log\,T$), surface gravity ($log\,g$), radius ($radius$), and luminosity ($log\,L$).

\begin{figure*}[t]
    \centering
    \includegraphics[width=\textwidth]{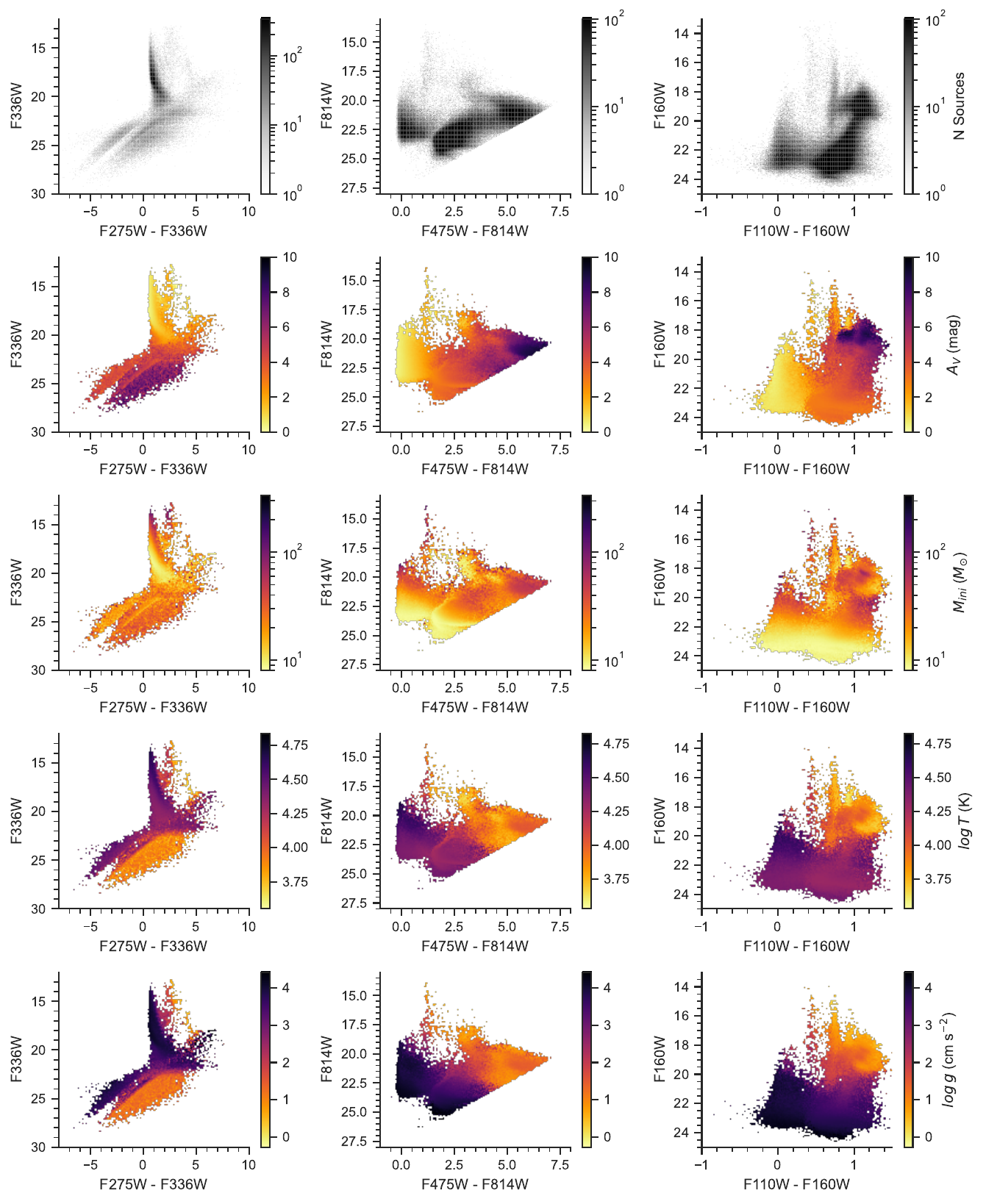}
    \caption{Top row: Density distribution of initial 467k massive star candidate sources in the catalog in NUV (left), visible (center), and NIR (right) color-magnitude diagrams (CMD). Lower rows: Binned median extinction ($A_{V,\,LOS}$), initial stellar mass ($M_{ini}$), effective temperature ($log\,T$), and surface gravity ($log\,g$) of candidate sources, derived from the BEAST.}
    \label{fig:full_cmd}
\end{figure*}

One advantage of the BEAST is the ability to incorporate prior knowledge about the underlying parameter distribution e.g. low-mass stars are far more likely to be observed than high-mass stars. To incorporate this information, a similarly-sized prior grid is constructed using the priors for each parameter. The range and resolution of each parameter determine the size of the grid. For each star, the BEAST reports the expectation value (\texttt{Exp}) along with the 16th, 50th, and 84th percentile of every parameter by marginalizing over the posterior probability distribution function (pPDF). A full pPDF is computed from the likelihood of every model SED multiplied by its prior value, where the likelihood is computed assuming a multi-variate Gaussian model, i.e. how does the measured flux of a star compare with the model? From these results, we can obtain the expectation value of the line of sight extinction ($A_{V,\,LOS}$) towards each individual star.
%How deeply embedded are massive stars, and does this depend on how intense the local star formation is, and/or the local ISM column density?

\section{Catalog}\label{sec:catalog}

We create a preliminary catalog of 467,412 massive star candidates by sub-selecting stars with BEAST fits that have expected initial mass estimates equal to or greater than 8 $M_{\odot}$, the typical mass limit above which stars are expected to undergo a Type II supernova explosion. We use initial mass estimates rather than current/actual mass estimates ($M_{act}$) since massive stars tend to lose a substantial fraction of their mass over the course of their lives due to stellar winds \citep{Castor1975, abbott1982}. %stars will eventually fuse enough iron in their cores to exceed the Chandrasekhar limit (1.4 $M_{\odot}$), triggering an energetic Type II core-collapsed supernova explosion. 
In Figure \ref{fig:full_cmd}, we plot the 2D density distribution (top), the median line-of-sight extinction ($A_{V,\,LOS}$), initial mass ($M_{ini}$), effective stellar temperatures ($log\,T$), and surface gravity ($log\,g$) derived from the BEAST for all massive star candidates $M_{\odot}$ in NUV (left), visible (middle), and NIR (right) CMDs. 

We find a large fraction of abnormal candidate sources in Figure \ref{fig:full_cmd}, most clearly seen in the second NIR CMD colored by $A_{V,\,LOS}$. These sources were fit by the BEAST to have line-of-sight extinction $\simeq 3.36$ mag, significantly higher than the median total column density extinction at the locations of the massive stars $\simeq 0.67$ mag. These sources are extremely dim in the UV and are characterized as massive stars with low effective temperature ($log\,T$) and high line-of-sight extinction ($A_{V,\,LOS}$).

Due to their abnormally high extinction estimates, we strongly suspect that the majority of the sources in the catalog are in fact asymptotic giant branch (AGB) stars. Non-stellar contaminants were originally removed from the PHAT catalog using sharpness and roundness cuts on the photometry, ensuring that the BEAST fit mostly stellar sources. However, Version 1.0 of the BEAST\footnote{https://github.com/BEAST-Fitting/beast/releases/tag/v1.0} used to process the PHAT photometry files did not include stellar models for AGB stars. These are evolved intermediate-mass stars (0.8-8 $M_{\odot}$) that produce considerable amounts of dust via dredge-ups during the final phases of their evolution. As a result, these stars often appear highly extinguished, not because of their galactic environment, but because of their dusty stellar winds. This creates a degeneracy in the BEAST where dusty intermediate-mass AGB stars end up being classified as highly-extinguished massive stars.

\subsection{Asymptotic Giant Branch Stars}\label{sec:agb}
% We corroborate this hypothesis by comparing our catalog with other AGB catalogs.
We compare our massive star candidate catalog to two AGB catalogs from \citet{goldman2021} and \citet{massey2021}. 
\citet{goldman2021} used NIR photometry from PHAT to produce a near-complete catalog of 346,623 Thermally-Pulsing AGB stars in the M31 PHAT footprint. \textit{Spitzer} IRAC observations were crossmatched to obtain further IR coverage and confirmation. Since this AGB catalog is directly derived from the PHAT data set, we want to compare our massive star catalog with additional AGB catalogs derived from independent measurements.
\citet{massey2021} identified 265,519 potential AGB contaminants while classifying red supergiants in M31. Through private correspondence, Dr.~Phillip Massey provided us with the M31 AGB data set used to create Figure 10 in their paper. The data set includes ICRS coordinates for each source, as well as photometry measurements in $K$ (2.2 $\mu m$) and $J-K$ bands, obtained with the wide-field NIR camera WIRCam on the 3.6 m Canada-France-Hawaii Telescope, which are redder than the PHAT NIR bands (F110W and F160W). Since these sources were observed independently from PHAT, we compare the spatial coordinates of the AGB candidates from \citet{massey2021} with the entire PHAT catalog, keeping matches with less than 0.9$^{\prime \prime}$ of separation (3-pixel radius aperture with 0.3$^{\prime \prime}$/pixel). In total, we find spatial matches to 104,002 sources in the PHAT catalog.

\begin{figure}
    \centering
    \includegraphics[width=\linewidth]{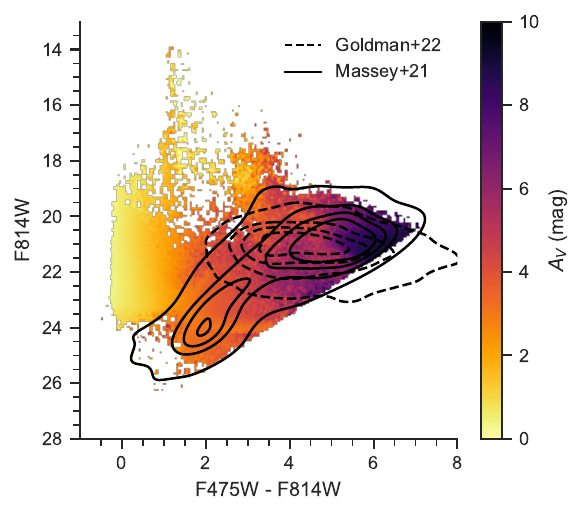}
    \caption{Visible CMD of candidate massive sources in the catalog, colored by $A_{V,\,LOS}$. Kernel density estimate contours of AGB candidate sources from \citet{massey2021} (solid) and \citet{goldman2021} (dashed) are overlaid on top. We suspect the highly extinguished stellar populations are in reality just misclassified AGB sources, based on their co-location on the CMD. %To remove potential AGB contaminants, we exclude all sources below the black line,  $F814W = -3\times(F475W-F814W) + 26$.
    }
    \label{fig:agb_cmd}
\end{figure}

In Figure \ref{fig:agb_cmd}, we compare the visible CMD of our massive star candidate catalog, colored by $A_{V,\,LOS}$, to the AGB catalogs from \citet{goldman2021} and \citet{massey2021}. We find that the AGB stars overlap with high-extinction sources from our catalog, indicating that the these sources are likely misclassified AGB stars. % We remove all these sources by making a color cut in the visible CMD. This removes 414,071 sources, the majority of the massive star candidates (. 
To effectively remove as many of the potential contaminant AGB sources as possible, we use the derived BEAST parameters to separate the two populations.

\begin{figure*}
    \centering
    \includegraphics[width=\textwidth]{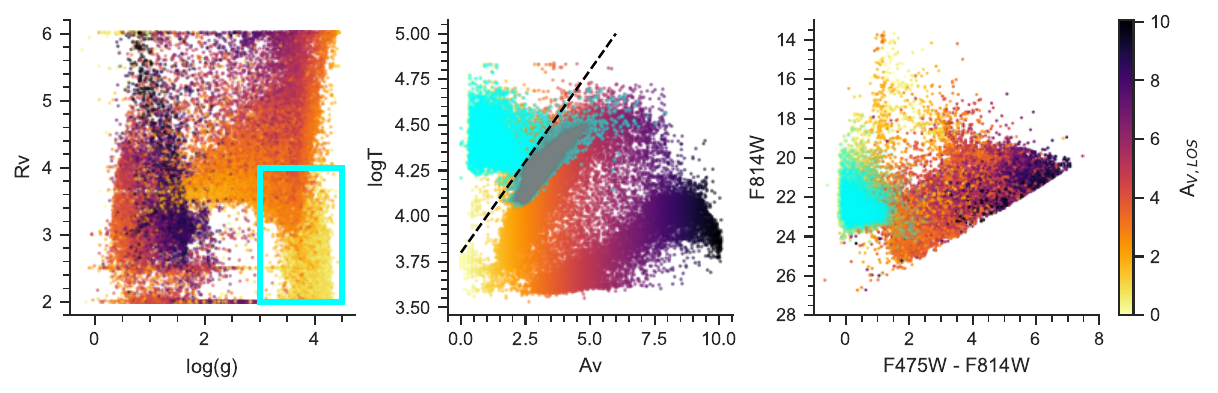}
    \caption{Left: BEAST-derived total-to-selective extinction ($R_V$) and stellar surface gravity ($log\,g$) measurements of all 467k PHAT sources originally identified as massive stars by the BEAST ($M_{ini} > 8 M_{\odot}$). We separate real massive stars from misclassified AGB stars by only selecting low-extinction sources with $log\,g > 3$ and $R_V< 4.0$, shown in the cyan box (n=139k). Middle: BEAST-derived extinction ($A_{V,\,LOS}$) and stellar temperature ($log \, T$) for the same catalog, with sources from the previous box selection shown in cyan. Our final catalog only consists of sources with high temperatures (n=50k), found above the black dashed line defined as $log\,T_{eff} = 0.2 A_V + 3.8$. Right: Visible CMD of the original catalog versus the 50k massive star candidates based on our selections. For clarity, only 10\% of the original catalog is shown.}
    \label{fig:beast_cuts}
\end{figure*}

In Figure \ref{fig:beast_cuts} (left), we show the BEAST-derived total-to-selective extinction ($R_V$) as a function of the derived stellar surface gravity ($log\,g$). By coloring the sources based on the extinction ($A_{V,\,LOS}$), we see that the high-extinction sources (i.e. misclassified AGB stars) span a wide area of parameter space while all low-extinction sources (i.e. most likely actual massive stars) are confined to a small range of $log\,g$ and $R_V$ values. To separate low-extinction sources from high-extinction contaminants, we select all sources with $log\,g$ between $3.0 - 4.5$, and $R_V$ between $2.0-4.0$ (cyan box, $n=139,133$ sources). This selection removes most of the high-extinction sources from our catalog, with the exception of a small population of low-$R_V$ contaminants that overlap with our low-extinction sources in the upper-left region of our selection. To separate these two overlapping populations, we use BEAST-derived extinction ($A_{V,\,LOS}$) and stellar temperature, ($log\,T$) to further sub-select the catalog (Figure \ref{fig:beast_cuts}, middle; cyan scatter). Low-extinction stars generally have a higher effective temperature than the misclassified AGB sources, making it possible to separate the two populations. We derive the following equation by eye to bisect the massive star candidates from the high-extinction AGB contaminants:

$$log\,T_{eff} > 0.2 A_{V,\,LOS} + 3.8$$

We test the effects of this cut  (Figure \ref{fig:beast_cuts} middle, dashed black line) in Section \ref{sec:variable_agb}. 

In Figure \ref{fig:beast_cuts} (right), we show the visible CMD (cyan) of the 49,916 selected massive stars above the dashed black line compared to the original catalog with 467k sources. We find that, by removing misclassified AGB stars using the BEAST-derived parameters, we are able to retain more reddened stars than if we had implemented a color cut on the photometry directly.  

\subsubsection{$A_V$ Detection Limits}\label{sec:av_limits}

There is a point at which a massive star will no longer be observable in all filters due to extinction lowering the apparent magnitude beyond the completeness limits. These completeness limits can be highly dependent on the source density in a region (using all PSF sources), where blending from neighboring stars limits the depth of the photometry. This is especially the case for IR filters in PHAT (F110W, F160W) where lower-mass stars are more readily detectable. In regions of low source density, F275W is often the limiting filter with a 50\% completeness limit at $\sim$24 mag, regardless of source density. A summary of the completeness limits for each band in PHAT can be found in Table 8 in \citep{williams2014}.

To test the limits of $A_V$ at which massive stars are no longer observable, we use PARSEC isochrone models\footnote{http://stev.oapd.inaf.it/cgi-bin/cmd} to generate a sample of massive stars SEDs of varying masses (8-30 $M_{\odot}$) and extinction (0.5 dex). All stars have an age of 5 Myrs and are at a distance of 752 kpc. Assuming these stars are in a low source density region, we find that an 8-M$_{\odot}$ star is no longer observable in F275W past $A_V \geq 2.5$ mag, whereas a 30-M$_{\odot}$ star is first no longer observable in F275W past $A_V \geq 4.5$ mag.

\subsubsection{Selection Effects}\label{sec:effects}

The youngest and most embedded massive stars cannot be observed, even with NIR coverage, and will therefore not appear in the catalog. However, as massive stars age and become less embedded, they eventually become detectable. This detection depends on their UV flux and initial mass, as discussed in Section \ref{sec:av_limits}. Since we are forced to utilize the BEAST-derived $A_{V,\,LOS}$ parameter in our quality cuts (Figure \ref{fig:beast_cuts}, middle), we need to ensure that this cut does not accidentally remove partially embedded sources that might otherwise be observed.

To investigate whether our quality cut generates a selection bias in our catalog, we examine how the distribution of $A_{V,\,LOS}$ changes as a function of stellar initial mass. The most massive stars can probe higher levels of obscuration and therefore sample a more complete distribution of $A_V$ across stages of embeddedness. Without selection effects, the shape of this $A_V$ distribution should repeat at all other mass ranges. However, we would expect the median $A_V$ to decrease at lower mass ranges since these stars cannot sample as high an extinction. If selection effects are present, we would observe an unnatural cutoff in the $A_V$ distribution from the quality cut removing high $A_V$ sources in that mass range.

\begin{figure}
    \centering
    \includegraphics[width=\linewidth]{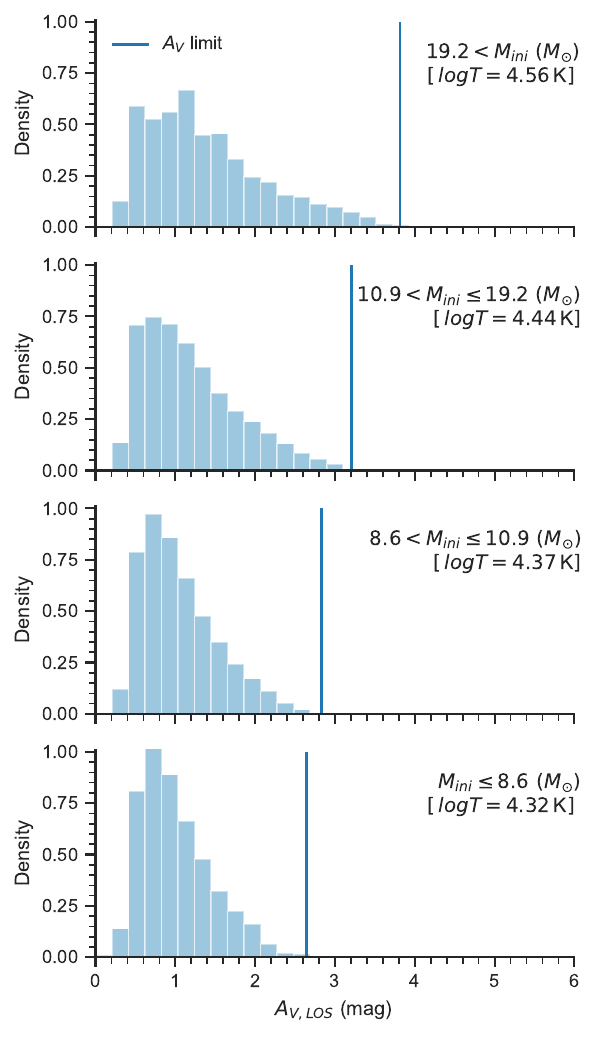}
    \caption{Distribution of BEAST-derived $A_{V,\,LOS}$ split by different mass bins spanning the 16th, 50th, and 84th quantiles. The corresponding $A_V$ limit imposed by the quality cuts is shown as a vertical line.}
    \label{fig:selection}
\end{figure}

In Figure \ref{fig:selection}, we plot the distribution of $A_{V,\,LOS}$ as a function of mass bins spanning the 16th (8.6 $M_{\odot}$), 50th (10.9 $M_{\odot}$), and 84th (19.2 $M_{\odot}$) quantiles. For each mass range, we compute the median effective temperature and calculate the corresponding $A_V$ limit based on the equation in Section \ref{sec:agb}. The distribution of $A_{V,\,LOS}$ in the highest mass bin follows a log-normal distribution, similar to the expected column density distribution of dust, as discussed in Section \ref{sec:Av25}. In all lower mass bins, we observe similar log-normal distributions, indicating that the lack of lower-mass massive stars at high $A_V$ is a product of observation effects due to increasing embeddedness, and not a product of the quality cuts we impose. Since the $A_V$ limit is consistently higher than the tail of the $A_V$ distribution in each mass range, it is safe to assume that the quality cut is not removing a significant number of highly-extinguished observable sources that could alter our subsequent analysis.

% Given these observational constraints, the quality cut on $logT$ and $A_{V,\,LOS}$ will not be removing any significant populations of extinguished sources that could alter our subsequent analysis. 

\subsubsection{Foreground Sources}

A minor additional advantage of BEAST-derived quality cuts is that we remove potential foreground stars. While foreground stars were ideally removed as part of the original photometric quality cuts, \citet{williams2014} expect foreground contamination in the PHAT catalog to be $<0.02\%$, based on the entire catalog of over 100 million sources. By modeling the expected magnitudes of foreground stars across the wavelengths observed with PHAT, they found that any foreground stars would form a vertical strip at $\sim$1 in F475W-F814W with a range from $14-28$ mag in F814W on a visible CMD \citep[see Figure 19 in][]{williams2014}. We observe a similar stripe of bright stars in the original catalog (colored), as seen in Figure \ref{fig:beast_cuts} (right), which have conveniently been omitted from the final catalog. These sources were originally classified as extremely bright evolved massive stars, since the BEAST assumes a single distance towards M31 when fitting. Fortunately, our quality cuts remove these sources from the catalog since our first selection (Figure \ref{fig:beast_cuts}, right) imposes a strict limit on the minimum stellar surface gravity ($log\,g \geq 3$). While this cut removes any actual evolved massive stars, it is hard to justify their inclusion since we have limited models of evolved massive stars, meaning that any parameter fits would probably have large uncertainties.

\subsection{Other Quality Cuts}\label{sec:cuts}

In this section, we describe and justify all additional quality cuts imposed on our catalog.

We suspect some of our sources to be duplicates due to their small spatial separation. The minimum pixel separation needed for DOLPHOT to resolve separate stars is set by the \texttt{Rcombine} parameter. For PHAT, \texttt{Rcombine}=$1.415$ pixels based on F814W photometry \citep[see Table 2,][]{williams2014}. Based on the resolution of UVIS (0.04$^{\prime \prime}$/pix), we calculate that the minimum possible separation between stars has to be at least 0.0566$^{\prime \prime}$ or $\sim$0.2 pc assuming a distance of 752 kpc to M31 \citep{riess2012}. We find 918 sources with neighbors closer than this distance. We omit these sources from our catalog since these sources are duplicates from overlapping PHAT bricks that the BEAST fit multiple times.

% \begin{figure}
%     \centering
%     \includegraphics[width=\linewidth]{./logg_cut.pdf}
%     % \includegraphics[width=\linewidth]{./chi2min_cut.pdf}
%     \caption{Color magnitude diagram (CMD) of massive star candidates colored by their surface gravity. We opt to exclude potential misclassified binary candidates by removing any stars with $log\,g <3$.}
%     \label{fig:logg_cmd}
% \end{figure}

% \subsubsection{Foreground and Binary Stars} 

% We find 5,089 sources in our catalog that we suspect to be misclassified binary systems or foreground sources due to their extreme brightness and low $log\,g$ estimates and exclude them from our catalog.

% \subsubsection{Duplicate Fits}

% \subsubsection{Poor Regional Extinction Fits}

Lastly, we omit 6,891 sources that reside in areas with low dust ($A_{V,\,25} \leq 0.5$) since we do not have good constraints on the regional extinction in those areas (see Section \ref{sec:Av25}). 

\begin{figure*}
    \centering
    \includegraphics[width=0.95 \linewidth]{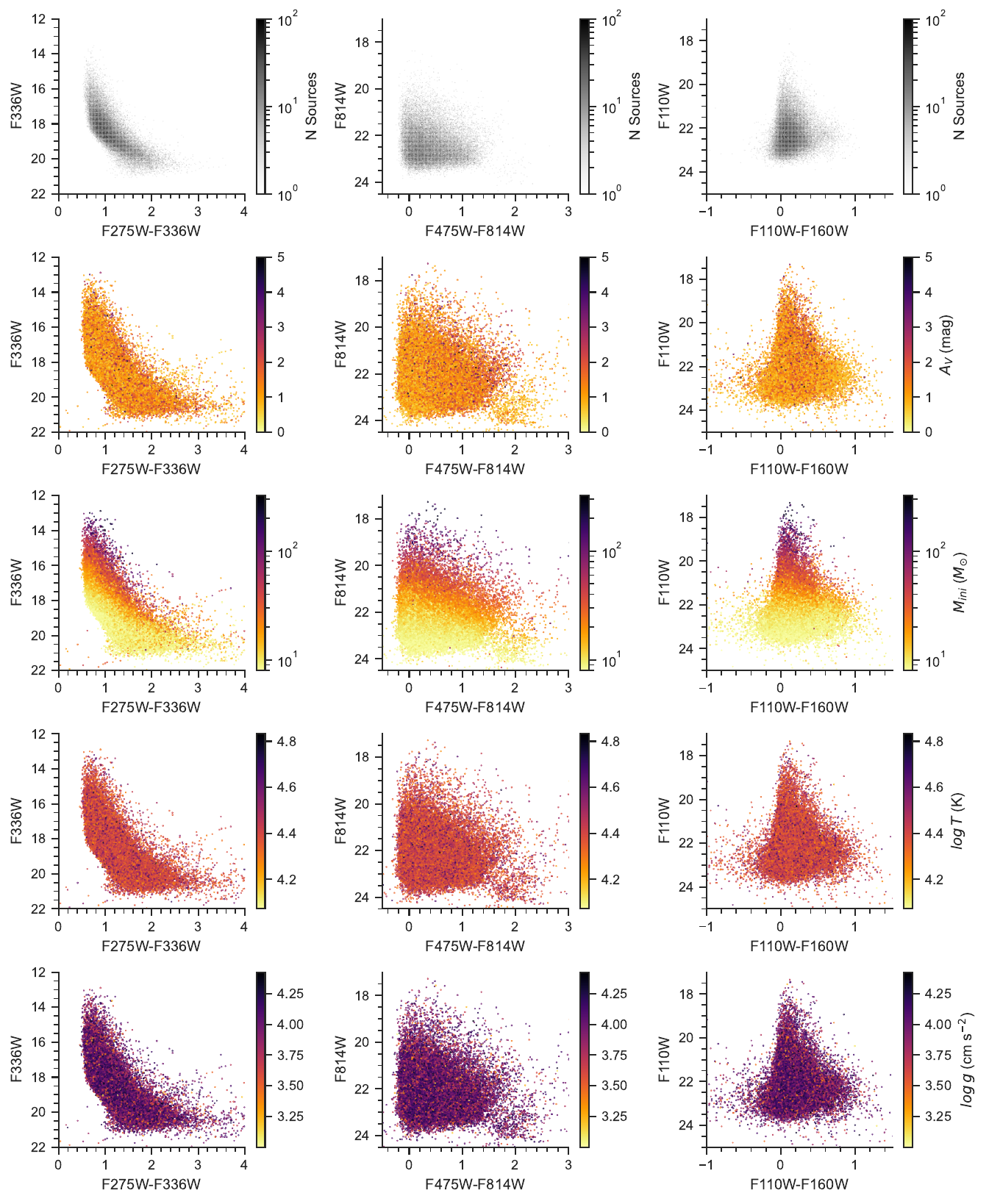}
    \caption{Top row: Density distribution of massive star candidates in the final catalog ($n=42,107$) in NUV (left), visible (center), and NIR (right) CMDs. Lower rows: Scatter plots of extinction ($A_{V,\,LOS}$), initial stellar mass ($M_{ini}$), effective temperature ($log\,T$), and surface gravity ($log\,g$) of final catalog sources, derived from the BEAST..}
    \label{fig:CMD_final}
\end{figure*}

\begin{figure*}
    \centering
    \includegraphics[width=0.95 \textwidth]{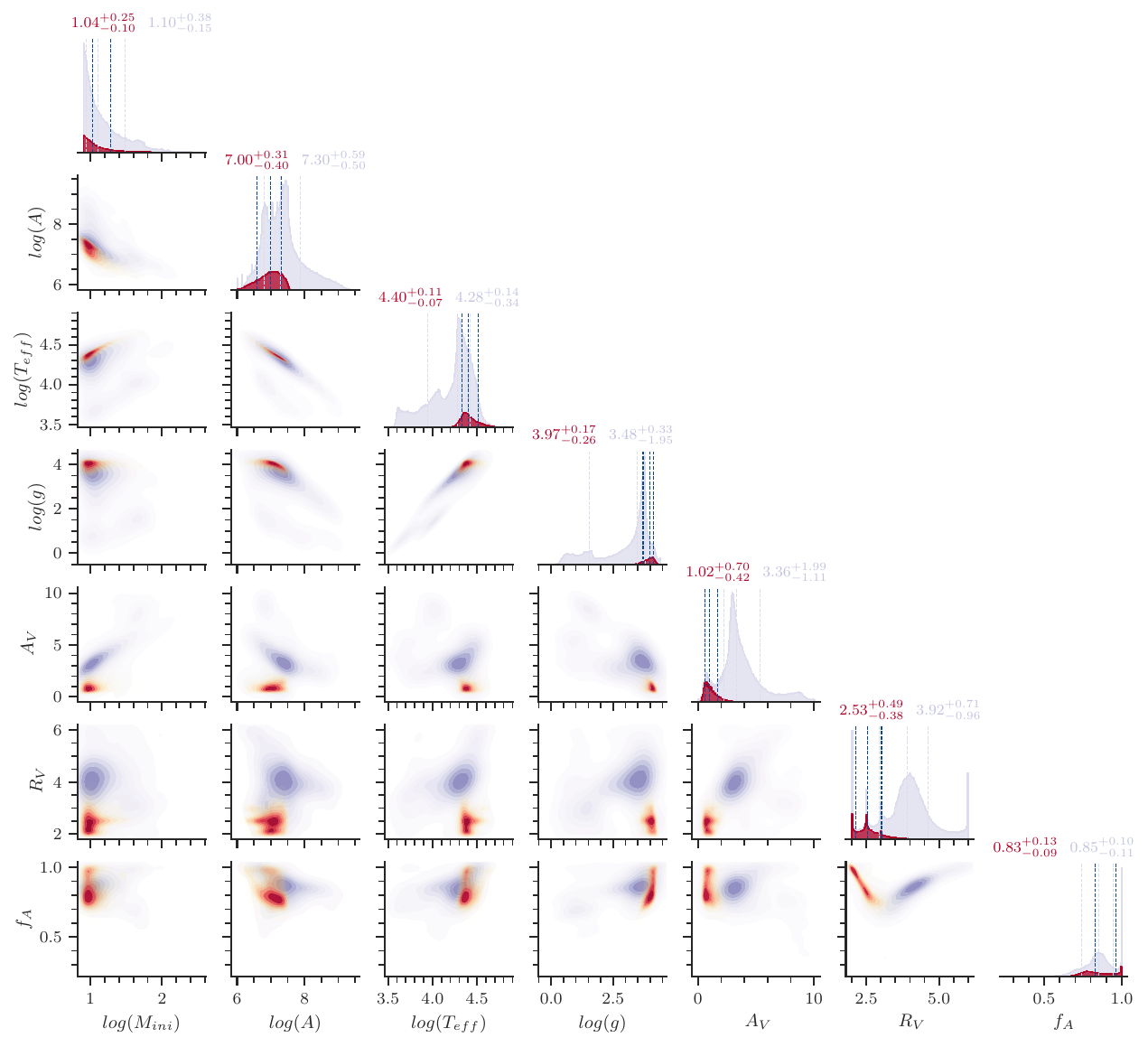}
    \caption{Kernel density estimate distributions of expectation values of stellar and dust BEAST parameters for the full untrimmed catalog (purple, n=467k) and our final catalog (red, n=42k). BEAST parameters: initial mass ($M_{ini}$), age ($log\,A$), temperature ($log\,T_{eff}$), surface gravity ($log\,g$), extinction ($A_{V,\,LOS}$), total-to-selective extinction ($R_V$), and SMC-like extinction curve ($f_A$), where \texttt{0} is SMC-like and \texttt{1} is MW-like.}
    \label{fig:beast_param_full}
\end{figure*}

\subsection{Final Catalog}\label{sec:final_catalog}
% section with final CMDs

Our final catalog is the result of the following selection process using the BEAST-derived stellar and dust expectation parameters (\texttt{Exp}):

\begin{enumerate}
    \item $M_{ini} \geq 8 M_{\odot}$ ($n = 467,412$)
    \item $log_{10}(g) \geq 3$ \& $R_V \leq 4$ ($n = 139,133$)
    \item $log_{10}(T_{eff}) > 0.2 A_V + 3.8$ ($n = 49,916$)
    \item $dist({\text{nearest neighbor})} \geq 0.0566^{\prime \prime}$ ($n=48,998$)
    \item $A_{V,\, 25} > 0.5 \text{ mag}$ ($n = 42,107$)
\end{enumerate}

The final catalog of massive star candidates consists of 42,107 sources. This catalog is available at MAST as a High Level Science Product\footnote{\url{https://archive.stsci.edu/hlsp/phathighmass/}} via \dataset[10.17909]{\doi{10.17909/z936-tm54}}. NUV, visible, and NIR CMDs are shown in Figure \ref{fig:CMD_final}. The catalog sources now span a far smaller range of magnitudes, especially in the NUV. 

Distributions of key BEAST parameters are shown in Figure \ref{fig:beast_param_full}.  Our final catalog (blue) spans a smaller range of age, temperature, surface gravity, and initial mass than the original catalog (red). In addition, the average extinction ($A_{V,\,LOS}$) is roughly half the original catalog and now spans a range that is in agreement with findings from \citetalias{dalcanton2015}. While most of the BEAST parameters display a smooth distribution, there are a couple of parameters, particularly $R_V$, that display multimodal spikes in their distribution. These spikes fall on the locations of the BEAST parameter gridding, an indication that the parameters are not resolved. Since $R_V$ is a secondary nuisance parameter not broadly used in this analysis, we do not consider this a point of concern.

\begin{figure*}
    \centering
    \includegraphics[width=\textwidth]{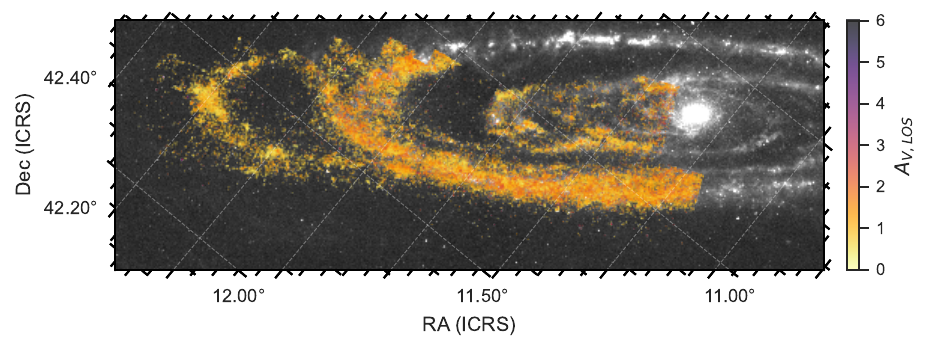}
    \includegraphics[width=\textwidth]{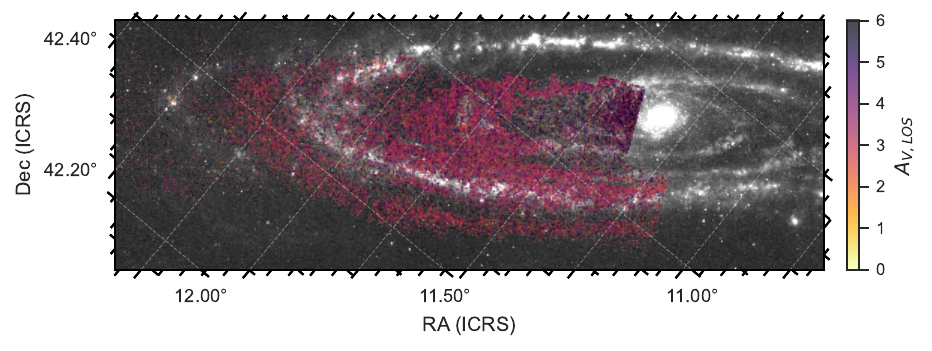}
    \caption{Top: Spatial distribution of massive star candidates (n=42,107) overlaid on a 24-micron dust map. The stars are colored by their BEAST-derived expected extinction. Bottom: Spatial distribution of omitted catalog sources overlaid on a 24-micron emission map. For clarity, only 10\% of the omitted sources (n=41,750) are shown.}
    \label{fig:ir_dust_map}
\end{figure*}

\subsection{IR Dust Emission Map} 

When dust grains are heated by nearby stars, they emit radiation in the infrared (IR). 24-micron radiation is a direct consequence of higher dust temperatures, stimulated by stellar UV photons \citep{calzetti2007}. Young massive stars are the primary contributors of UV photons, meaning that the presence of massive stars should result in an increase of 24-micron radiation. In Figure \ref{fig:ir_dust_map} (top), we compare the spatial distribution of our 42,107 massive star candidates against 24-micron from the Multiband Imaging Photometer for Spitzer \citep[MIPS,][]{gordon2006} and find that the stars in our catalog closely trace the hot dust. By comparison, the AGB sources we omit from the catalog using our selection process (bottom) tend to reside between the dust rings, another indication that these sources are not actual massive stars, but rather, a misclassified older stellar population homogeneously distributed across the disk.

\section{Comparing with Spectroscopically Classified Massive Stars}\label{sec:valid}

In this section, we compare our massive star catalog with spectroscopically classified massive stars. Based on previous observations for the Local Group Galaxy Survey (LGGS), \citet{massey2016} obtained spectroscopic follow-up observations of 1,895 stars in M31 and M33 using the Hectospec spectrograph on the MMT Observatory. They identified 64 O-type stars and 321 B-type stars in M31. We compare our massive star candidate catalog with the spectroscopically-identified massive stars from \cite{massey2016}. 

\begin{figure}[h]
    \centering
    \includegraphics[width=\linewidth]{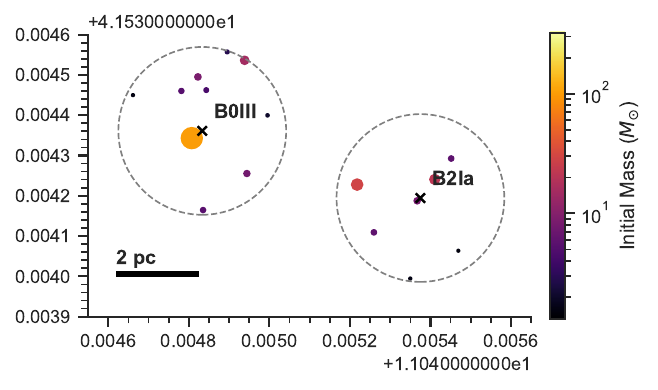}
    \caption{Two spectroscopically-identified B-type stars (black crosses), with all possible matches from PHAT within a diameter of 1.5", colored and sized by BEAST-derived initial mass estimates.}
    \label{fig:massey_example}
\end{figure}

We match each spectroscopic OB star to all sources in the PHAT BEAST catalog within a distance of 0.75$^{\prime \prime}$, based on the 1.5$^{\prime \prime}$ diameter of the Hectospec. As illustrated in Figure \ref{fig:massey_example}, each OB star has dozens of sources within the search radius, however, most of these sources are too faint or too red to produce an OB-type spectrum. We match each source in the spectroscopic catalog to the brightest F475W star that is within the search radius. We find suitable BEAST-fit matches for all 158 spectroscopically classified OB stars that lie within the PHAT observational footprint. 

\begin{figure}
    \centering
    \includegraphics[width=\linewidth]{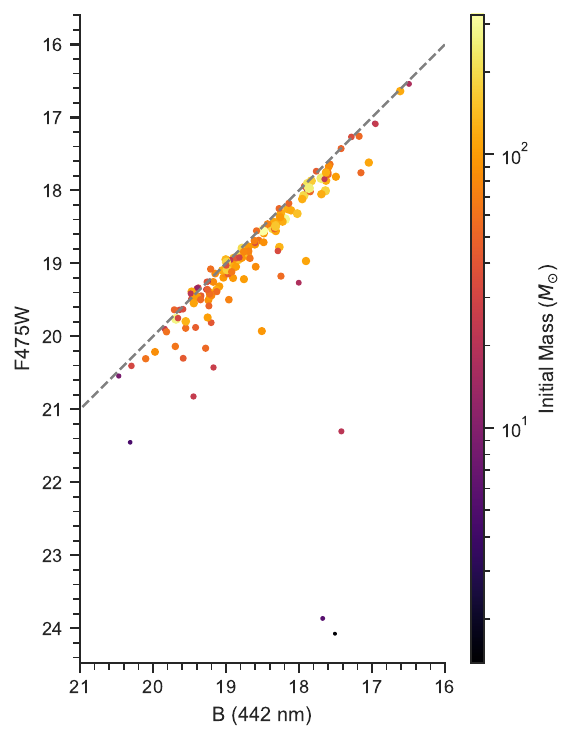}
    \caption{Comparison of visible magnitudes for the best photometric matches from the PHAT catalog (HST WFC3 F475W) to the spectroscopic OB sources from \citet{massey2016}. B-band magnitudes are derived from preliminary photometric observations from LGGS.}
    \label{fig:massey_best}
\end{figure}

Each spectroscopic OB star has photometric B and V-band measurements from the preliminary LGGS investigation. We compare the B-band magnitude (442 nm) with HST WFC3 F475W magnitudes of our PHAT sources in Figure \ref{fig:massey_best} and find strong agreement between the two measurements for most of the sources. We find only three sources with significant differences ($>1.5$ mag) between their F475W and B-band measurements. Two of these sources, however, are missing photometric UV coverage from F336W, and the other has two massive stars within the Hectospec fiber diameter ($M_{ini} = 21$ and $28 M_{\odot}$). We note that BEAST fits were originally run on any PHAT sources with detections in at least four bands, however, 

\begin{figure*}
    \centering
    \includegraphics[width=\textwidth]{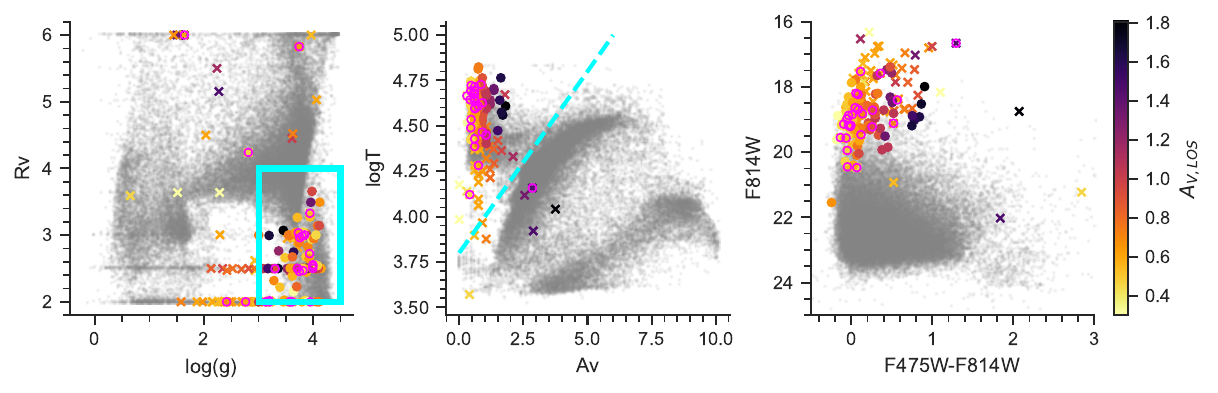}
    \caption{Left: BEAST-derived total-to-selective extinction ($R_V$) and stellar surface gravity ($log\,g$) measurements of the 158 PHAT sources matched with the spectroscopic OB stars \citep{massey2016}, colored by $A_V$. Sources marked with \texttt{X} would have been removed by massive star selection cuts ($n=71$, see Section \ref{sec:final_catalog} for details), and all sources in regions with low extinction ($n=25$, $A_{V, 25} < 0.5$ mag) are circled in magenta. The untrimmed catalog ($n=467k$) is plotted in grey for reference and our catalog selection cuts are indicated in cyan. Middle: BEAST-derived extinction ($A_V$) and stellar temperature ($logT$) for the same catalogs. Right: Visible CMD of the same catalogs. For clarity, only 10\% of the untrimmed catalog is shown across all panels.  
    }
    \label{fig:massey_cmd}
\end{figure*}

Although most of the 158 spectroscopic OB stars within the PHAT footprint have matches with massive stars in the PHAT catalog, only 87 OB stars are included in our final catalog. In total, 71 sources are excluded from our final catalog, of which, 43 sources are removed due to low values of $log\, g$, 14 sources are removed due to low values of $R_V$, and 25 sources are removed due to being located in regions with low $A_{V,\,25}$ \textbf{(Section \ref{sec:final_catalog})}. Note that these individual values do not add up to the total number of omitted sources since some sources are excluded in multiple quality cuts. Figure \ref{fig:massey_cmd} shows the location of the OB matches across the different quality cuts. Any sources not found in the final stellar catalog are marked with an \texttt{X} and all sources in low $A_{V,\,25}$ are circled in magenta. 

As shown in Figure \ref{fig:massey_cmd}, most of the spectroscopically matched sources are incredibly bright compared to the rest of the sources in the massive star catalog, which could be an indication that these sources are actually binary stars we are not able to resolve with HST. Based on theories of star formation, a large fraction of massive stars ($\sim 70\%$) are expected to exist in binary systems \citep{sana2012, neugent2021}. Binary stars (interacting and non-interacting) were not implemented in Version 1.0 of the BEAST, meaning that, if binary sources were not resolvable with HST, then the BEAST would proceed to fit the combined flux of the two sources as a single object. This is not an issue if the secondary star is much fainter than the primary star since the BEAST will essentially end up fitting the flux of the primary star. However, if the secondary star provides a significant contribution of flux relative to the primary star in any given band, then their combined fluxes could exceed the fluxes modeled in the BEAST grid. In this scenario, the best-fit model would by default be the brightest model i.e. an evolved massive star with low surface gravity and a poor-quality of fit. These sources would have been removed from our massive star catalog due to our quality cuts which impose a strict limit on the minimum surface gravity ($log\,g \geq 3)$ in order to remove misfit sources.

The fact that all 158 spectroscopically-observed OB stars have photometric matches in PHAT is excellent confirmation that PHAT does not suffer from any observational effects that would limit the detection of bright massive stars i.e. if the stars had been too bright and saturated their PSF. In addition, almost all of these stars were matched to PHAT sources that had been fit with $M_{ini} \geq 8$, showing that the BEAST is capable of characterizing massive stars. The fact that only 87 spectroscopic OB stars would have ended up in our catalog is entirely due to the quality cuts we apply to limit contaminants. Future versions of the BEAST will incorporate improved AGB models, meaning that fewer quality cuts will be needed to tease out potential contaminants and ensure a high level of purity for future catalogs.

\subsection{Other Massive Star Catalogs}

\begin{figure}
    \centering
    \includegraphics[width=\linewidth]{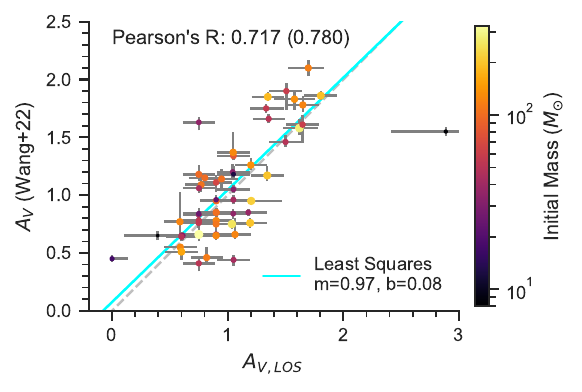}
    \caption{Extinction estimates and uncertainties from \citet{wang2022} compared to BEAST-derived line-of-sight extinction from \citetalias[][]{gordon2016}. We plot the one-to-one correlation in dashed grey. We compute the Pearson's product-moment correlation coefficient between the two extinct measurements for all data ($R=0.717$). Omitting the right-most high-uncertainty data point at [$2.9, 1.6$], we recompute the Pearson's correlation coefficient ($R=0.780$) and compute the least squares fit, shown in cyan.}
    \label{fig:wang22}
\end{figure}

Using spectra and photometry from the LGGS survey, \citet{wang2022} constructed multiband SEDs for 140 OB-type stars by incorporating observation from PHAT \citep{dalcanton2012, williams2014}, the Panoramic Survey Telescope and Rapid Response System Release 1 \citep[Pan-STARRS1,][]{hodapp2004, chambers2016}, the United Kingdom Infrared Telescope \citep[UKIRT,][]{lawrence2007}, the Ultraviolet and Optical Telescope \citep[Swift/UVOT,][]{roming2005, yershov2015}, and the XMM-Newton Serendipitous Ultraviolet Source Survey \citep[XMM-SUSS,][]{mason2001, page2019}. They then combined TLusty \citep{Hubeny2017} and ATLAS9 \citep{castelli2003} stellar atmosphere models with dust model extinction curves to model extinguished stellar SEDs across a wide range of wavelengths. Using Markov Chain Monte Carlo (MCMC) analysis, they fit the observed data to the modeled SEDs, giving them parameter estimates for all their fitting parameters ($\alpha$, $log\,T$, $log\,g$, $A_V$) and derived parameters ($E(B-V)$, $R_V$), where $\alpha$ is defined as the dust size distribution in the dust model which shapes the extinction curve.

Since their catalog spans the entire disk of M31, only 58 of their 140 sources lie within the PHAT region. We find matches to all 58 sources in our catalog. % They found a median $A_V$ $\approx$ 1 mag, with most extinction values ($\sim$90\%) being under 1.8 mag.
In Figure \ref{fig:wang22}, we compare SED-derived $A_V$ measurements from \citet{wang2022} to line-of-sight extinction estimates from the BEAST ($A_{V,\,LOS}$) and find strong agreement between the two measurements. Both measurements use photometry from PHAT to construct their SEDs, however, \citet{wang2022} incorporated data covering a larger range of wavelengths which, in theory, should improve their ability to constrain a good model fit to their SEDs. The fact that these extinction measurements were derived using a different method from the BEAST shows that the correlation between these two measurements is not a result of their data and method being similar, but rather a result of the resolution they probe within the ISM. Overall, the results from \citet{wang2022} corroborate the line-of-sight extinction measurements we obtained with the BEAST.

\section{Results}\label{sec:results}

\begin{figure}
    \centering    
    \includegraphics[width=\linewidth]{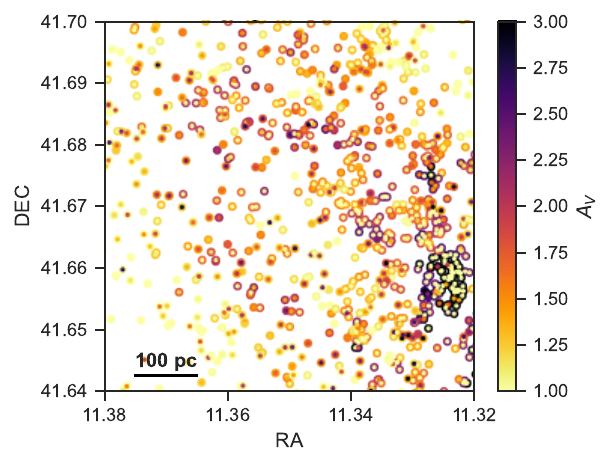}
    \caption{Map of catalog sources in a dense stellar region of M31, with the $A_V$ range limited to 1-3 mag. Points are color-coded both by the local mean extinction in 25-pc pixels from \citetalias{dalcanton2015} ($A_{V,\, 25}$), indicated by the color of the outer ring of each point, and by the line-of-sight extinction to each individual star ($A_{V,\, LOS}$), indicated by the color of the center of each point.  Differences between the inner and outer regions of a single point indicate that a star is viewed with less or more extinction than the typical extinction in that local region.}
    \label{fig:av_visual}
\end{figure}

Through visual inspection alone, we discover interesting trends in extinction measurements. In Figure \ref{fig:av_visual}, we compare the line-of-sight extinction measurements ($A_{V,\, LOS}$, center) from the BEAST with the regional 25-pc average extinction ($A_{V,\, 25}$, edge) from \citetalias{dalcanton2015}, in a region of M31 with a high density of massive stars. We identify associations of massive stars in regions with large amounts of dust (high $A_{V,\,25}$) with low line-of-sight extinction ($A_{V,\, LOS}$) towards the stars (lower right). These sources could be massive stars embedded in dense molecular clouds which are able to clear their nearby natal ISM through stellar feedback. However, there are also massive stars outside of these associations that exhibit high $A_{V,\, LOS}$ despite being in less dusty regions. These stars could be forming in-situ in the inter-arm regions of M31 in spurs small enough to be unresolved at 25-pc scales. To test these scenarios, we first classify stars as a continuous function of the local density of massive stars and then examine trends (and differences) between the line-of-sight and total column extinctions as a function of stellar density.

\subsection{Calculating Massive Stellar Densities} 

To quantify the vast range of stellar environments in M31, we calculate the underlying density distribution of massive stars. Rather than calculating the source density of massive stars, which suffers from discrete results at small length scales, we smooth the probability density of each star using Kernel Density Estimation (KDE) to approximate each star as a Gaussian kernel with some variance, i.e. bandwidth ($k$). This allows us to quantify the density of massive stars around any given massive star by summing the influence of all the surrounding massive stars with the following equation:

$$M = \sum e^{-D^2/2k^2}$$

where $D$ is the distance from one massive star to every other massive star and $k$ is the bandwidth or length scale of influence.

\begin{figure}
    \centering
    \includegraphics[width=\linewidth]{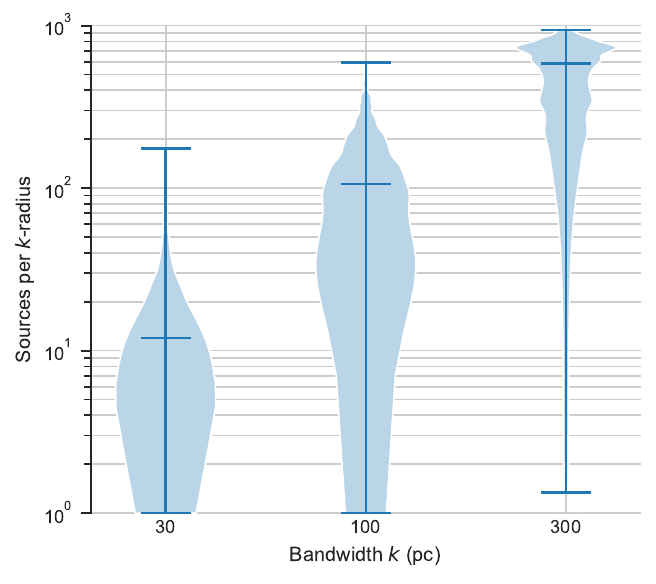}
    \caption{Stellar density ranges using 30, 100, and 300 pc as the bandwidth for the KDE. Medians are marked.}
    \label{fig:kde_ls}
\end{figure}

\begin{figure}
    \centering
    \includegraphics[width=\linewidth]{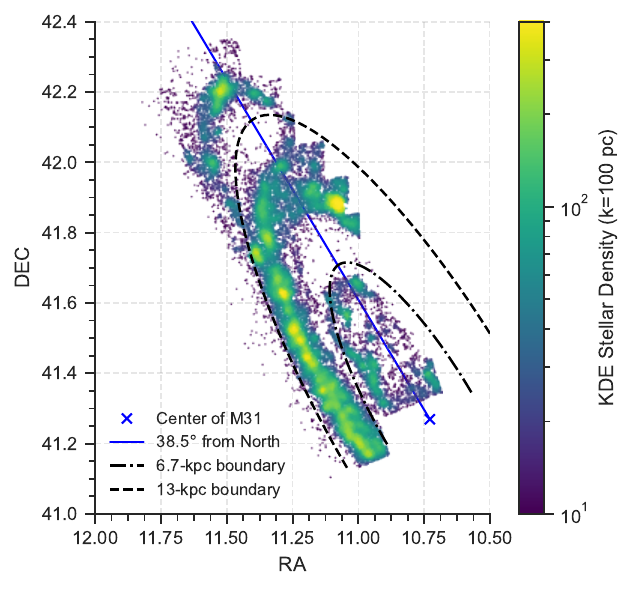}
    \caption{KDE map of stars using 100-pc bandwidth. Dashed lines show the boundaries between the 5, 10, and 15-kpc rings. Radial distances for massive stars were calculated assuming a major axis of 38.5$^{\circ}$ from North (up), an inclination of 74$^{\circ}$, and a distance of 752 kpc to M31. }
    \label{fig:kde}
\end{figure} 

% \begin{figure}
%     \centering
%     \includegraphics[width=\linewidth]{./inclination.pdf}
%     \caption{Radial distance of massive stars from the center of M31 assuming a major axis of 38.5$^{\circ}$ from North, an inclination of 74$^{\circ}$, and a distance of 752 kpc to M31. Dashed lines show the boundaries between the 5, 10, and 15-kpc rings.}
%     \label{fig:inclination}
% \end{figure}

We initially varied the bandwidth $k$ from 30 to 300 pc. These scales were motivated by the physical scales of molecular clouds, where globules can span as little as a few pcs, while molecular complexes can span several hundred pc \citep{Chevance2023}. However, as shown in Figure \ref{fig:kde_ls}, we find that the dynamic range of stellar densities is severely skewed at bandwidths of 30 and 300 pc, so we opt to use a bandwidth of 100 pc. 

We sample the underlying stellar density map by summing the probabilities at the location of every massive star, giving us a quantitative metric for how clustered of a region each massive star is in. Figure \ref{fig:kde} shows the spatial distribution of the catalog sources colored by their stellar densities. %To quantify the effect of bandwidth on the underlying stellar density maps and results, see Appendix \ref{app:kde_apdx}.

We bin the resultant stellar density distribution into 10 bins on a logarithmic scale and calculate the mean for both the line-of-sight ($A_{V,\, LOS}$) and regional ($A_{V,\, 25}$) extinction. We use block bootstrapping to quantify the standard error uncertainties. First, we split the entire map into 10,000 blocks, 100x100 (each block being 46.2$^{\prime \prime}$/44.7$^{\prime \prime}$ in RA/DEC), and select all the blocks that contain sources. Then, we sample these blocks with replacement $n$ times, where $n$ is the number of blocks that contain sources, to construct a new stellar catalog. We opt to bootstrap our uncertainties using blocks rather than individual data points so the probability of sampling sparser regions of massive stars was comparable to sampling dense regions of massive stars. Had we chosen to bootstrap individual points, the probability of sampling massive stars in dense regions would have been greater than in sparser regions. We repeat this sampling 100 times\footnote{In general, our results converge after 10 samples.}, calculating the mean line-of-sight and regional extinction per stellar density bin each time, before calculating the final mean and standard deviations of all the sampled average. Since the standard deviation is calculated using the means of the samples, we note that our final measurement of uncertainty is the standard error of the catalog. 

\subsubsection{Deprojecting the Stellar Catalog} \label{sec:deproj}

M31 is highly inclined along our line-of-sight (i=74$^{\circ}$) so, to measure stellar densities in physical units, we have to account for projection effects\footnote{The PHAT region is technically further away than the northwestern half of M31's disk. If this is hard to visualize, try rotating an image of M31 135$^{\circ}$ clockwise so that the PHAT region is above the northwestern half. This was suggested by Dr. Christopher Clark, who pointed out that humans usually have an easier time perceiving surfaces from the top down due to how we interact with most objects in everyday life.}. To do this, we rotate the catalog sources to align with the major axis, 38.5$^{\circ}$ from North, centered at RA=10.6847929$^{\circ}$ and DEC=41.2690650$^{\circ}$, and de-incline the sources by 74$^{\circ}$ based on their distance from the major axis \citepalias{dalcanton2015}. Then, we convert their angular distance from the center of M31 into kilo-parsecs (kpc), assuming a distance of 752 kpc to M31 \citep{riess2012}.

\begin{figure}
    \centering
    \includegraphics[width=\linewidth]{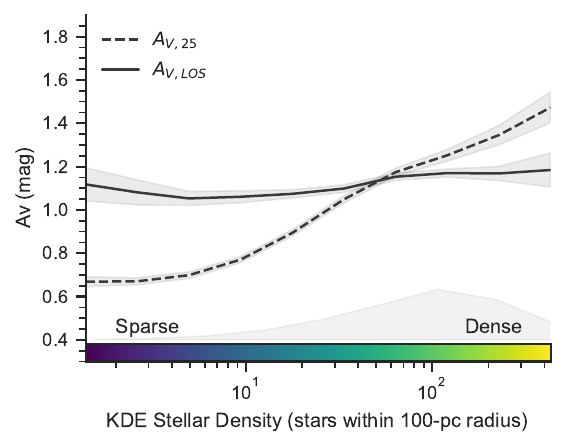}
    \caption{Regional $A_{V,\, 25}$ (dashed) and line-of-sight $A_{V,\, LOS}$ extinction (solid) binned by KDE stellar density. We use block bootstrapping (100x100 blocks, reps=100) to calculate the standard error (shaded). A scaled histogram of the stellar density distribution is plotted along the x-axis.}
    \label{fig:main_cumulative}
\end{figure}

\subsection{Stellar Density Variation}
% description of plot
We plot the resulting averages between line-of-sight extinction ($A_{V,\, LOS}$) and regional extinction ($A_{V,\, 25}$) as a function of stellar density in Figure \ref{fig:main_cumulative}. We find that $A_{V,\, 25}$ generally increases monotonically with stellar density. This is consistent with our understanding that massive stars are more likely to form in areas with significant amounts of dusty star-forming gas. We note that the flattened slope at lower stellar densities is likely due to the fact that we omit massive stars in areas with low regional extinction ($A_{V,\, 25} < 0.5$ mag) due to our quality cuts. 

Massive stars exhibit similar line-of-sight extinction regardless of their region in the galaxy, as seen by the flat distribution of $A_{V,\,LOS}$. The fact that $A_{V,\, LOS}$ is constant regardless of stellar density implies that there is the same amount of dust in the immediate environment of massive stars, regardless of their location in the galaxy. We also note that $A_{V,\, LOS}$ is less than $A_{V,\, 25}$ in regions of high stellar density, whereas $A_{V,\, LOS}$ is greater than $A_{V,\, 25}$ in regions of low stellar density. 

\subsection{Influence of Quality Cuts}\label{sec:variable_agb}

To test what sort of influence our quality cuts (Section \ref{sec:final_catalog}) could have had on our results, we reconstruct the catalog with variable quality cuts. 

\subsubsection{Low Regional Extinction}\label{sec:variable_av}

First, we test whether excluding sources in regions with low regional extinction had any significant effect on $A_{V,\, LOS}$ results. We reconstruct the catalog, including the 6,891 sources originally excluded in regions with low regional extinction ($A_{V,\, 25}<0.5$), recalculate the KDE stellar densities, and resample the $A_{V,\, LOS}$. We forego analyzing $A_{V,\, 25}$ since our catalog now includes sources with unreliable $A_{V,\, 25}$ measurements. 

\begin{figure}
    \centering
    \includegraphics[width=\linewidth]{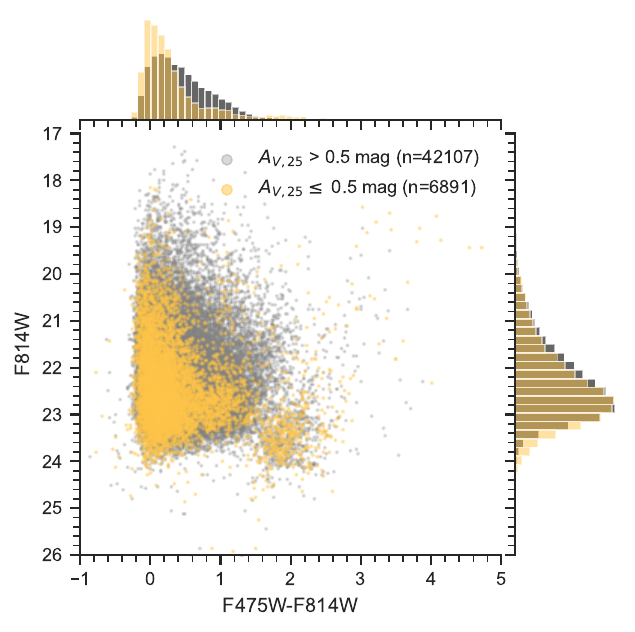}
    \hspace*{-1.8cm}\includegraphics[width=0.85\linewidth]{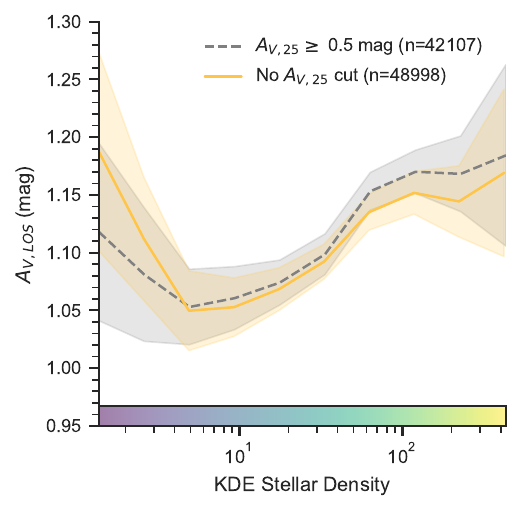}
    \caption{Top: CMD of original catalog (grey) and the sources excluded in regions with $A_{V,\, 25}<0.5$ (orange). Marginalized normalized histograms of both axes are plotted along the top and side. Bottom: $A_{V,\, LOS}$ as a function of KDE stellar density for the original catalog and the catalog without a quality cut on sources with $A_{V,\, 25}<0.5$. The only difference between the two catalogs is the omission of the fifth quality cut i.e. no stars are remove in areas with low $A_{V,\, 25}<0.5$. All the other quality cuts stay the same. Note that the y-axis scale is much smaller than the other result plots.}
    \label{fig:variable_av}
\end{figure}

In Figure \ref{fig:variable_av} (top), we plot the visible CMD of the massive star catalog (grey, $n=42,107$) versus the sources originally excluded from low extinction regions (orange, $n=6,891$). We find that the excluded sources have similar magnitude distributions as the catalog sources, indicating that the excluded sources do not suffer from any observational biases. The excluded sources are generally bluer than the catalog sources, which is to be expected given that these sources are located in areas with less regional extinction. 

In Figure \ref{fig:variable_av} (bottom), we plot $A_{V,\, LOS}$ as a function of stellar density for the massive star catalog (grey, $n=42,107$) and the massive star catalog \textit{plus} the excluding sources (orange, $n=48,998$). We find that excluding sources in regions with low extinction has no significant effect on $A_{V,\, LOS}$ as a function of stellar density. The biggest difference in $A_{V,\, LOS}$ can be seen in regions of low stellar density, which is not surprising since these regions also generally have the lowest average regional extinction. However, the slight increase in $A_{V,\, LOS}$ is well within the standard error of the measurements, indicating that excluding these sources from our analysis has no discernable effect on our results.

\begin{figure}
    \centering
    \includegraphics[width=\linewidth]{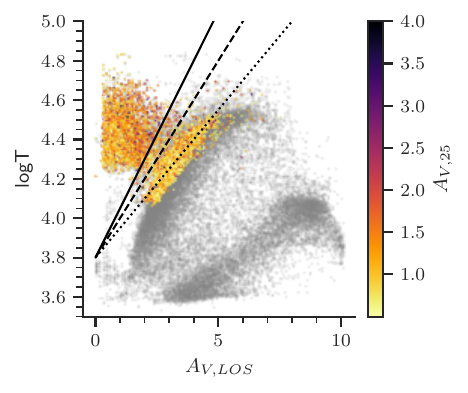}
    \caption{BEAST-derived extinction ($A_{V,\,LOS}$) and stellar temperature ($log\,T$) for all sources originally selected using the first ($M_{ini} \geq 8 M_{\odot}$) and second ($log \,g \geq 3$ \& $R_V \leq 4$) quality cuts, colored by the regional extinction ($A_{V,\,25}$). The entire initial catalog of 417k candidate sources is shown in grey in the background. The slope of the third quality cut is altered to create a conservative catalog (solid: $m=0.25$), the original catalog (dashed: $m=0.2$), and a liberal catalog (dotted: $m=0.15$), after which all subsequent quality cuts are applied as normal i.e. removing sources with duplicate neighbors and/or sources in low $A_{V,\,25}$ regions.} %The visible CMD of stellar catalogs using different slopes for the AGB cuts. The original catalog of untrimmed stars with initial mass estimates greater than 8 is shown for reference in grey. }
    \label{fig:variable_cuts}
\end{figure}

\begin{figure*}
    \centering
    \includegraphics[width=\textwidth]{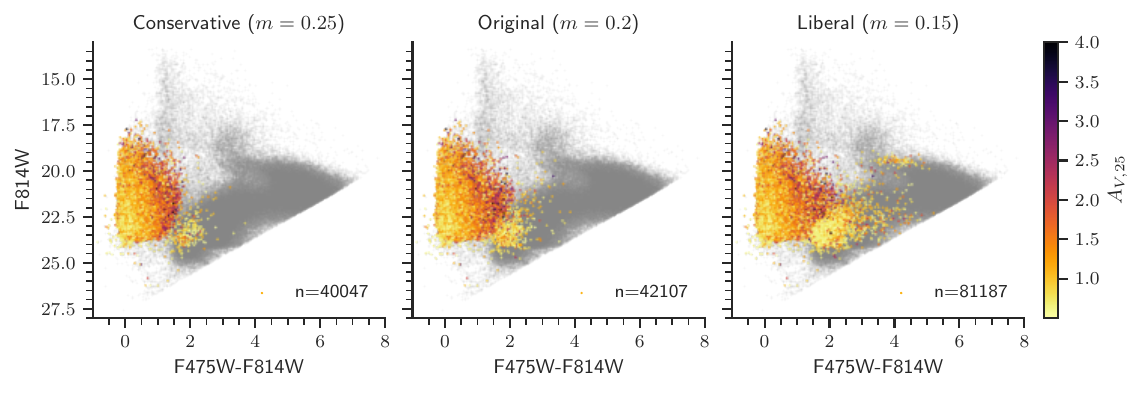}
    \caption{Visible CMD of the conservative (solid: $m=0.25$), original (dashed: $m=0.2$), and liberal (dotted: $m=0.15$) quality cut catalogs, colored by the regional extinction ($A_{V,\,25}$).}
    \label{fig:variable_cuts2}
\end{figure*}

\subsubsection{Influence of AGB Sources} 

Next, we test how AGB contaminants might influence our results by modulating the original slope of Step 3 in our quality cuts. We reconstruct our catalogs with the same quality cuts described in Section \ref{sec:final_catalog}, now changing the slope of our third cut ($log\,T > 0.2 A_V +3.8$) to produce three new catalogs: $m=0.25$ for a more conservative cut (n=40,047), and $m=0.15$ for a more liberal cut (n=81,187). Figure \ref{fig:variable_cuts} illustrates these cuts. With these two new catalogs, we reconstruct the KDE stellar density maps and recalculate the mean $A_{V,\, LOS}$ and $A_{V,\, 25}$ as a function of stellar density. 

In Figure \ref{fig:variable_cuts2}, we show the resultant distribution of sources in each catalog on a visible CMD, colored by the regional extinction measurements ($A_{V,\, 25}$). In general, we find that redder massive stars come from areas with higher amounts of regional extinction ($A_{V,\, 25}$). This is not surprising given that, if a star is embedded in a greater column density of dust, then its photometry will appear dimmer at shorter wavelengths, resulting in a redder color. One major exception to this trend is the presence of a low extinction clump located at F475W-F814W $\approx 2.5$. This clump of sources grows in number when we relax our quality cut on $A_{V,\,LOS}$ v. $log\,T$ and shrinks when we restrict it. Based on this trend, these sources must be located near the boundary of our quality cut which could be an indication that they are a potential source of contamination. Upon further investigation, we find that these sources are generally located in lower stellar density regions, making up $\sim25\%$ of the sources in our sparsest stellar density bin (KDE stellar density $\approx$ 1). Since these sources were not omitted by our original quality cuts, we flag them in our final catalog as potential contaminants.

% \begin{figure*}
%     \centering
%     \includegraphics[width=\linewidth]{./variable_cuts2.pdf}
%     \label{fig:variable_cmd_top}
%     \caption{The visible CMD of stellar catalogs using different slopes for the AGB cuts. The original catalog of untrimmed stars with initial mass estimates greater than 8 is shown for reference in grey. }
% \end{figure*}

\begin{figure}
    \centering
    \includegraphics[width=\linewidth]{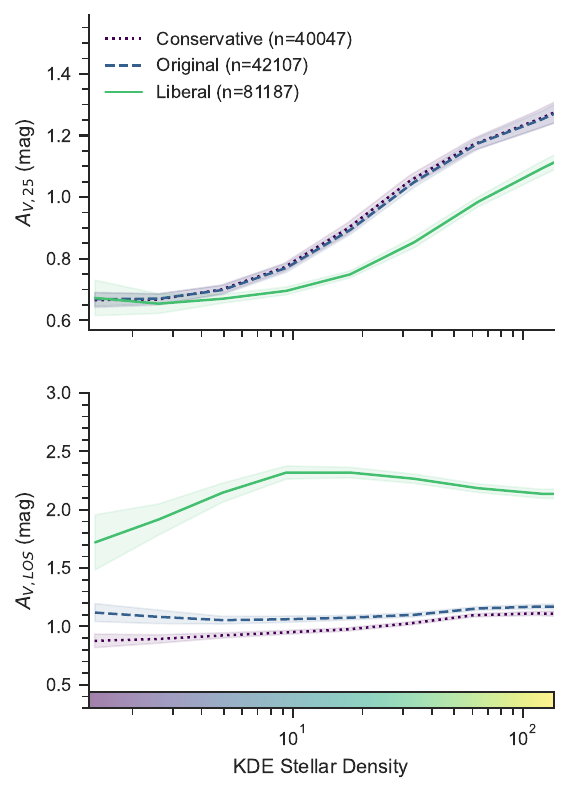}
    \caption{Block bootstrapped averages and uncertainties for $A_{V,\, 25}$ (top) and $A_{V,\, LOS}$ (bottom) as a function of stellar density using the different selection criteria shown in Figure \ref{fig:variable_cuts}.}
    \label{fig:variable_cmd}
\end{figure}

In Figure \ref{fig:variable_cmd}, we plot the block bootstrapped averages and uncertainties for $A_{V,\, 25}$ and $A_{V,\, LOS}$ as a function of quality cuts. We do not find any significant differences in $A_{V,\, 25}$ when we make our quality cut more conservative, %modulate the slope of our cut to $m=-4$ or $m=-2$, 
however, the more liberal quality cut catalog has lower $A_{V,\, 25}$ at high stellar densities. This is likely due to the fact that, by including more contaminating sources, the average stellar density increases. In addition, since AGB sources are more homogeneously distributed across the disk of the galaxy (Figure \ref{fig:ir_dust_map}, bottom), contaminating sources will be located in areas with much lower $A_{V,\, 25}$, effectively lowering the average regional extinction. %The fact that the more conservative catalog ($m=0.25$) shows similar trends in $A_{V,\, 25}$ as the original catalog is an indication that the two catalogs sample the same dusty environments. This provides further support that our original catalog is predominantly composed of actual massive stars.

%Our catalog is likely missing these highly embedded young massive stars still enshrouded in their natal ISM. Had these stars been included in our analysis, we predict the average $A_{V,\, LOS}$ would have increased uniformly across all stellar densities, assuming star formation across all galactic environments. $A_{V,\, 25}$ should remain the same though since these stars sample the same regions.

By relaxing our quality cut ($m=0.15$) to include AGB sources, we observed a significant increase in $A_{V,\, LOS}$ by almost one magnitude. This is to be expected, given that the AGB sources we removed from the catalog all had much higher $A_{V,\, LOS}$, $\sim 3.36$ mag compared to $0.99$ mag for catalog sources. By comparison, the conservative catalog remained almost identical to the original catalog. Since $A_{V,\, LOS}$ remains the same when restricting our quality cuts, we conclude that our catalog is not significantly contaminated by AGB sources. %We know that the group of low extinction sources comprises a significant fraction of sources at low stellar densities (KDE stellar density $\approx$ 1). When increasing the number of contaminating sources from this group in our catalog, we observe an increase in the average $A_{V,\, LOS}$ across all stellar densities. Therefore, we deduce that contaminating sources artificially inflate the average $A_{V,\, LOS}$ at all stellar densities.  %However, even with this inflation, the average $A_{V,\, LOS}$ still remains relatively constant across all stellar density environments, the only major difference being that $A_{V,\, LOS}$ now exhibits a slight positive correlation with stellar density. 

\subsection{Comparison with Gas Tracers}\label{sec:gas}

There are full-sky maps of atomic and molecular gas covering the PHAT region that approach the resolution of the regional total column density extinction map (25 pc) from \citetalias{dalcanton2015}. %(Table \ref{tab:gas_tracers}). 
In the following two sections, we compare how these gas tracers vary as a function of massive star stellar density.

% \begin{table*}[]
% \caption{Overview of Gas Tracers}\label{tab:gas_tracers}

% \begin{tabular}{lllll}
% \hline
% Author & Instrument & Tracer & Resolution \\ 
% \hline
% \citet{smith2021}\footnote{HARP and SCUBA-2 High Resolution Terahertz Andromeda Galaxy Survey (HASHTAG)}  & James Clerk Maxwell Telescope & Predicted CO (J=3-2) & 8$^{\prime \prime}$/30 pc\footnote{Assuming a distance of 652 kpc \citep{riess2012}} \\
% \citet{koch2021} & Karl G. Jansky VLA & HI 21-cm & 10$^{\prime \prime}$/32 pc 
                
% \end{tabular}

% \end{table*} 

\subsubsection{CO}

Carbon monoxide (CO) is a molecular gas that predominantly exists in cold ($\sim$5 K) regions of galaxies, often serving as a tracer for total molecular gas. In particular, CO has a dipole moment that makes certain rotational transitions, such as J=1-0 at $\lambda$=2.6 mm, easily observed at low temperatures relative to other species. It should be noted, however, that CO can often underestimate the total molecular gas mass since CO is more easily destroyed by UV photodissociation than other molecules such as molecular hydrogen, $H_2$ \citep{wolfire2010, bolatto2013}.

As the nearest star-forming disk galaxy to the Milky Way, M31's advantage in distance becomes a disadvantage in coverage given its angular size. The last complete CO map of M31 was created by \citet{nieten2006}, who used the IRAM 30-m telescope to construct a CO(J=1-0) map of the entire M31 galaxy at a resolution of 23$^{\prime \prime}$ (85 pc along the major axis). Since then, partial CO maps with higher resolution (5$^{\prime \prime}$.5/20 pc) have been created with the CARMA Interferometer \citep{calduprimo2016}, but these maps are not publicly available. As a compromise, we use CO (J=3-2)\footnote{ CO (J=3-2) is thought to trace hotter gas than CO (J=1-0), making it a slightly worse predictor of total molecular gas.} predictions from the HARP and SCUBA-2 High-Resolution Terahertz Andromeda Galaxy Survey \citep[HASHTAG, ][]{smith2021} to see how molecular gas varies as a function of stellar density. 

HASHTAG used the James Clerk Maxwell telescope to obtain ground-based submillimeter images of M31 at 450 $\mu$m and 850 $\mu$m which measure emission from dust \citep{eales2012}. One possible contaminant of the 850 $\mu$m image is the CO (J=3-2) transition line which, at $\lambda$=867 $\mu$m, can constitute anywhere from 1\% to 41\% of 850 $\mu$m emission in nearby galaxies \citep{smith2019}. By using other high-resolution CO surveys such as the CARMA survey \citep{calduprimo2016}, they were able to use linear models to predict CO(J=3-2) fluxes at a resolution of 8$^{\prime \prime}$ (30 pc).

\begin{figure}
    \centering
    \includegraphics[width=\linewidth]{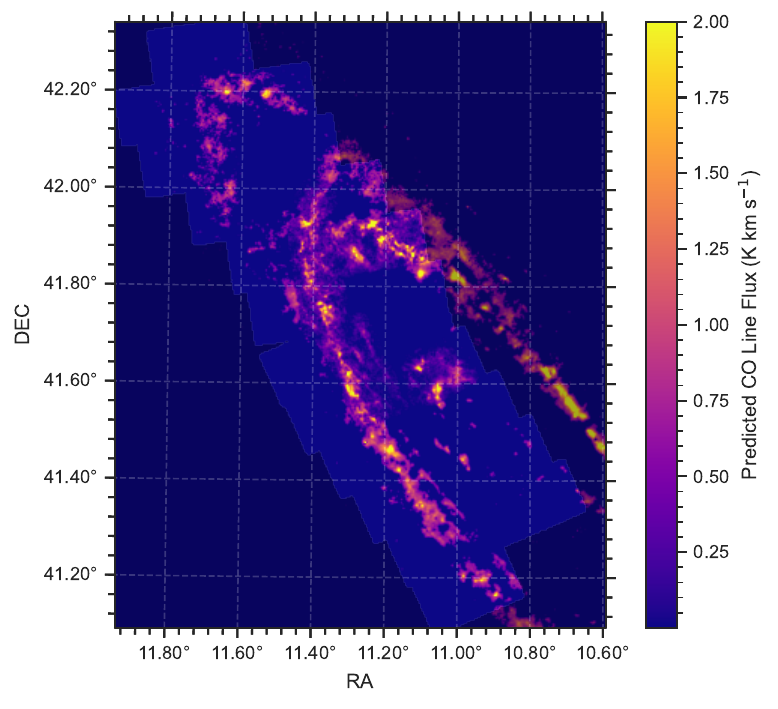}
    \hspace*{-1.0cm}\includegraphics[width=0.9\linewidth]{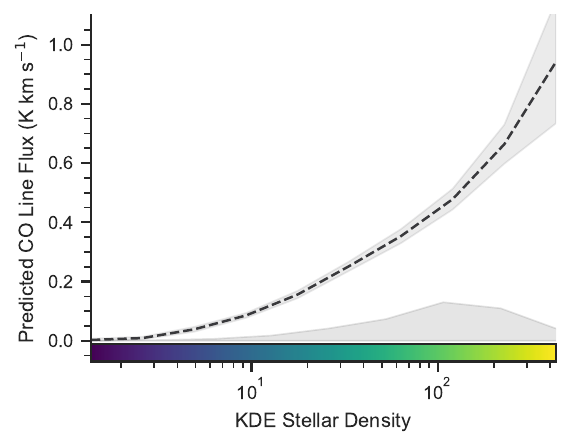}
    \caption{Top: Spatial distribution of the predicted CO(J=3-2) flux from the HASHTAG survey \citep{smith2021} compared to the PHAT footprint (dark mask). Bottom: Average predicted CO(J=3-2) flux as a function of stellar density derived from the massive star candidate catalog.}
    \label{fig:co32_coverage}
\end{figure}

To measure how CO varies as a function of stellar density, we follow a similar methodology as our results. We start by obtaining the predicted CO (J=3-2) flux at the location of each massive star candidate. We then bin the average CO emission as a function of stellar density, again using block bootstrapping to calculate the uncertainties. 

Figure \ref{fig:co32_coverage} shows how CO varies as a function of stellar density. We find that the CO monotonically increases as a function of stellar density, much like $A_{V,\, 25}$ in Figure \ref{fig:main_cumulative}. Most notably, CO seems to be almost entirely absent at lower stellar densities at the resolution of the HASHTAG survey. Higher-resolution molecular gas observations are needed to confirm whether molecular gas is entirely absent in these lower-density interarm regions or whether molecular clouds exist in spurs that are not resolvable at 30-pc scales. 

\subsubsection{HI}

Atomic hydrogen (HI) is the most common form of atomic gas and is broadly observable across most galactic disks via its emission at 21 cm. While generally hotter and less dense than molecular clouds, the temperature and density of HI can vary by several orders of magnitude depending on whether it is in equilibrium at a cold neutral medium (CNM) phase or warm neutral medium (WNM) phase. At both phases, the dust is thought to be well-mixed with the gas. As such, we would expect HI to vary similarly to $A_{V,\, 25}$ as a function of stellar density. 

\begin{figure}
    \centering
    \includegraphics[width=\linewidth]{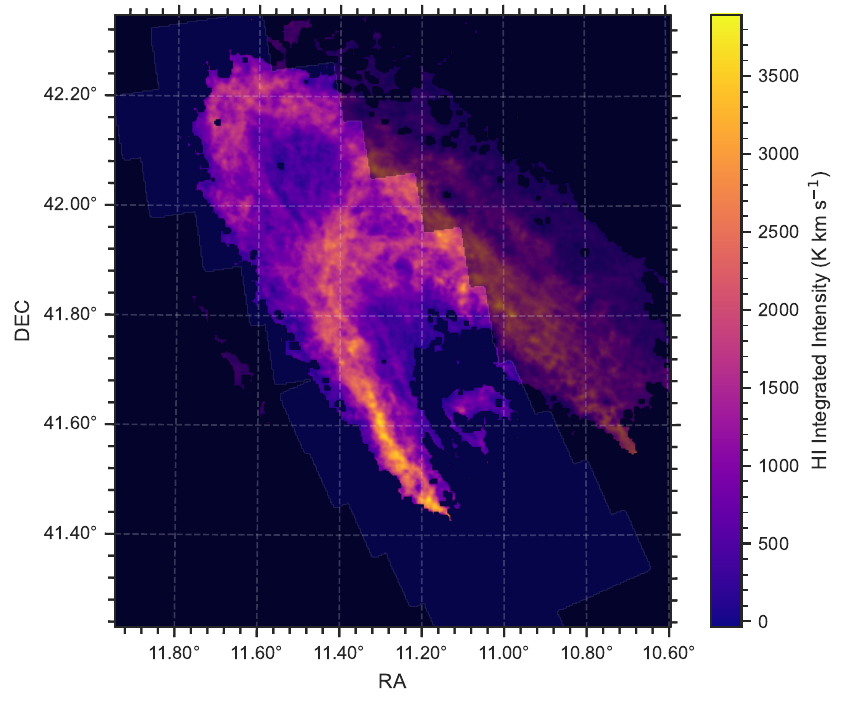}
    \hspace*{-1.4cm}\includegraphics[width=0.9\linewidth]{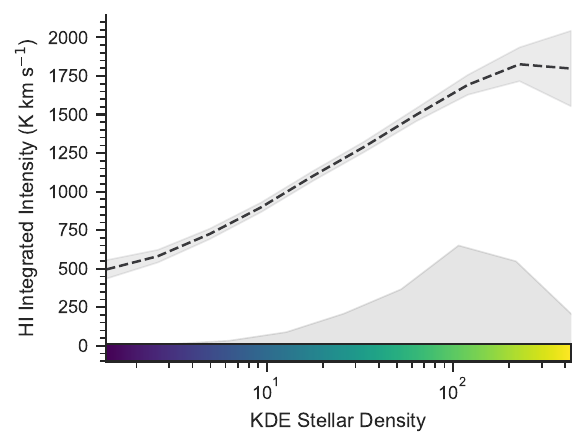}
    \caption{Top: Spatial coverage of HI 21-cm emission from \citet{koch2021} compared to the PHAT footprint (dark mask). Bottom: Average  HI 21-cm integrated intensity emission a function of stellar density derived from the massive star candidate catalog.}
    \label{fig:HI_coverage}
\end{figure}

\citet{koch2021} used the Karl G. Jansky Very Large Array (VLA)  to obtain HI observations of M31 at a spatial resolution of 10$^{\prime \prime}$ ($\sim$32 pc along the major axis). We use the integrated intensity measurements to estimate how HI varies as a function of stellar density using the same modifications as CO. We find that the average HI integrated intensity emission also decreases as a function of stellar density (Figure \ref{fig:HI_coverage}), although the rate is less than the predicted CO (J=3-2). The rate is most similar to the regional total column density extinction, which both decrease to $\sim$50\% of their maximum flux when going from higher clustered regions (KDE $\approx 10^2$) to sparse interarm regions (KDE $\approx 10^0$). At high stellar densities, the average HI integrated intensity flattens and uncertainties increase, likely due to atomic hydrogen transitioning to molecular hydrogen. 

\section{Discussion}\label{sec:disc}

The majority of star formation occurs within giant molecular clouds spanning 50-100 pc, meaning that, at 25-pc resolution ($A_{V,\, 25}$), we should be resolving most ISM structures. As such, we would expect the line-of-sight extinction ($A_{V,\, LOS}$) to sample extinction at 25-pc resolution and therefore exhibit similar trends as a function of stellar density. There are, of course, caveats to this assumption. If we assume that massive stars are emitting feedback, then $A_{V,\, LOS}$ should decrease at all stellar densities. Alternatively, if the ISM has filamentary substructure at scales smaller than 25-pc that massive stars trace, then $A_{V,\, LOS}$ should generally be higher than $A_{V,\, 25}$. However, from first principles, we expect the two measurement extinctions to exhibit similar trends across stellar densities.

% interpertation of plot: star formation
It is therefore notable that our findings show the average line-of-sight extinction ($A_{V,\, LOS}$) towards massive stars in M31 remains relatively constant across all stellar density environments despite the average regional extinction ($A_{V,\, 25}$) increasing monotonically with stellar density (Figure \ref{fig:main_cumulative}). In our attempts to understand these findings, we consider and discuss several potential sources of influence. %return to our initial research questions discussed in Section \ref{sec:intro}: Are massive stars observed in less dense regions of galaxies the result of runaway kinematics or were they formed there? 

\subsection{What are the effects of runaway stars?} 

We first consider the scenario that the existence of massive stars in regions of low stellar density are the result of runaway kinematics. Most massive stars are born in binaries and multiple systems \citep{mason2009, Almeida2017, moe2017}, and as such have a higher probability of absorbing momentum from their companions and being kicked from clusters \citep{renzo2019}. While the majority of massive stars exhibit peculiar velocities on the order of a couple of km s$^{-1}$ \citep{swiggum2021, galli2019}, $\sim$10\% of massive stars are observed to be runaway stars, with velocity dispersion +30 km s$^{-1}$ \citep{blaauw1961}. If these stars are born in the outskirts of the star-forming rings and are ejected at an angle tangential to the rings, then some of these runaway stars have the potential to migrate into low stellar density regions, even within their limited lifespans.

If all massive stars in regions of low KDE stellar density were the result of migration, we would expect the average line-of-sight extinction to be similar to the average regional extinction, given that the massive stars would be sampling the log-normal column density distribution of the region. Instead, we find that the line-of-sight extinction remains the same across all KDE stellar densities, meaning the majority of massive stars at low KDE stellar densities in the catalog cannot be the result of runaway kinematics.

In addition, we find 5\% of massive star candidates in our catalog have proper motion measurements from Gaia DR3 (n=2,221). Assuming a distance of 752 kpc to M31, $<$1\% (n=21) of these stars have proper motions greater than 30 km s$^{-1}$. For this preliminary analysis, we forgo correcting for any systemic tangential velocity since the velocity vector of M31 is primarily radial toward the Milky Way \citep{vandermarel2012}.

\subsection{Could massive stars be forming in low density environments?}

Next, we consider the possibility that massive stars are forming in situ in regions with low total column density extinction. The majority of star formation occurs within giant molecular clouds spanning 50-100 pc, producing a stellar population that roughly traces the initial-mass function \citep[IMF, ][]{mckee2007, bolatto2008, hughes2010}. This poses the question: Are massive stars able to form in low column density environments?%How small can molecular clouds be and still produce massive stars? 

% limited theoretical angle
%We usually think of massive star formation as being the inherent result of sampling from the IMF, meaning that, during the formation of one massive star, hundreds of lower-mass stars should also have formed. 
Massive star formation is theorized to occur either via competitive accretion, where large proto-stellar cores are created by merging smaller proto-stellar cores \citep{bonnell2006, peters2010}, or monolith collapse, where proto-stellar cores continue to accrete diffuse material until they become massive \citep{mckee2003}. The main difference between these two formation mechanisms is the mass distribution of lower-mass stars that accompany the formation of a massive star, shifting the peak of the IMF. However, both formation mechanisms are thought to predominantly occur in the densest parts of molecular clouds where the gravitational potential is the largest. As a result, there is currently limited evidence that either of these formation mechanisms could explain massive star formation in low column density regions. %As a result, massive stars are most often observed as members of gravitationally-bound stellar clusters with hundreds of smaller mass stars \citep{ladalada2003}.

%Massive stars have also been observed as members of unbound stellar systems such as OB associations. OB associations do not seem to follow the same spatial distribution as gravitationally-bound stellar clusters. Instead, they tend to trace the underlying distribution of gas, leading to a more hierarchical structure \citep{gouliermis2018}, evidence that OB associations are not only the product of disrupted stellar clusters.

% observational angle
While theoretical support of in-situ massive star formation in low density environments is limited, observational evidence of isolated massive star formation has been found in other galaxies like the Small and Large Magellanic Clouds \citep{oey2013, harada2019}. To confirm that massive field stars formed in situ in extremely sparse star-forming environments, a couple of criteria need to be met. Firstly, the stars cannot have large peculiar velocities relative to the surrounding stars or ISM. Secondly, there needs to be some evidence of star formation nearby, either via the detection of CO or an HII region. In the Small Magellanic Cloud, \citet{oey2013} detected 14 massive field stars that showed strong evidence for having formed in situ based on their proximity to HII regions. In the Large Magellanic Cloud, \citet{harada2019} used follow-up observations of CO from the Mopra telescope (resolution $\sim 11$ pc) to confirm the presence of compact molecular clouds near high-mass isolated young stellar objects (YSOs). These studies provide concrete evidence of in-situ massive star formation in low-extinction environments. 

% transition to theory of length-scales of molecular clouds
Observations show that massive stars can be formed in areas with low star formation activity. However, scientific theory suggests that these regions cannot be low density. %Although most giant molecular clouds and molecular complexes are usually observed spanning [X] pc, the ISM is thought to exhibit some amount of self-similarity, meaning that, across a wide range of resolutions, the column density distribution of the ISM will usually follow a power-law slope or log-normal distribution \citep{gouliermis2018}. 
In this paper, we primarily use the average extinction value obtained from regional 25-pc total column measurements. However, we expect that the column density of dust follows a log-normal distribution meaning that, even if a region has a low average extinction value, it could still contain higher density structures on smaller scales. Thus, we predict that massive stars in sparse regions could be forming from molecular clouds that are smaller than what we can detect with column density observations at 25-pc resolution. %Although most existing observations of gas tracers in M31 do not probe spatial resolution smaller than 25 pc, 

The simplest way to confirm that massive stars are forming in situ in low stellar density would have been to compare our results with observations of gas tracers at resolutions smaller than 25 pc. Unfortunately, publicly available observations of gas tracers in M31 are limited in resolution, as was shown in Section \ref{sec:gas}, meaning that these observations are not able to probe gas structures at resolutions smaller than 25 pc.
%Full-sky gas tracers all show similar trends as the average regional extinction, however, their resolutions are not high enough to compare with the line-of-sight extinction we derive from the stellar SEDs. 
As such, high-resolution follow-up observations are needed in order to fully validate the line-of-sight extinction measurements obtained with the BEAST and confirm our hypothesis that massive stars are forming in situ in low stellar density environments. Potential follow-up strategies are discussed in Section \ref{sec:followup}. %If these areas in interarm regions of galaxies reach sufficiently high column densities, then these spurs should still be able to form massive stars, albeit far fewer than giant molecular clouds. 

% In Section \ref{sec:future}, we discuss potential follow-up strategies for confirming our results.

% In addition, new high-resolution observations from JWST show that polycyclic aromatic hydrocarbons (PAHs) - small dust grains that trace the ISM - are pervasive across nearby star-forming galaxies observed by PHANGS-JWST \citep{thilker2023ApJ...944L..13T}. These dust grains form filamentary structures, both within and between spiral arms, hinting that gas structures are present in inter-arm regions. 

%There are several possible explanations for the counterintuitive results in Figure \ref{fig:main_cumulative}. 

%As such, our results support the picture that massive stars in low-density regions, such as inter-arm spurs, have similar small-scale interstellar environments (and thereby birth cloud properties) as massive stars in high-density spiral arms.

% Given that the resolution of the CO(J=3-2) (30 pc) is nearly the same size as the total column density extinction (25 pc), it is interesting to note that the total column extinction is much more similar to the HI integrated intensity emission (80 pc) despite the larger resolution. This could be an indication that the relative density of atomic gas is scale-invariant i.e. does not change as a function of resolution. 

\subsection{How are pre-supernova feedback mechanisms from massive stars altering the ISM?}\label{sec:feedback}

% From Figure \ref{fig:main_cumulative}, we find that the line-of-sight extinction around massive stars ($A_{V,\, LOS}$) is significantly lower than the average regional extinction ($A_{V,\, 25}$) at high stellar densities inside star-forming rings. One potential explanation for this discrepancy between line-of-sight and regional extinction is that pre-supernova feedback from massive stars is clearing the natal ISM surrounding them while leaving the rest of the molecular clouds undisturbed. 
Massive stars produce a variety of pre-supernovae feedback mechanisms, most notably stellar winds and photoionization which ionize and shock surrounding gas, creating HII regions and high-temperature bubbles in the ISM \citep{McLeod2021, Chevance2023}. Since feedback effects are cumulative over time, we would expect older and less massive stars to have lower $A_{V,\, LOS}$ than their younger counterparts as pre-supernova feedback clears the natal ISM surrounding them. In this section, we investigate how line-of-sight extinction varies as a function of age, using initial mass estimates as a proxy, to assess whether we find evidence of pre-supernova feedback around massive stars. 

Uncertainties of BEAST parameters are inherently a product of their original grid spacing or resolution. To characterize as many stars as possible, parameters like stellar age span a logarithmic scale, ranging from $10^6$ to $10^{10}$ years with a resolution of 0.05 dex \citepalias[see Table 1,][]{gordon2016}. The most massive of stars only live a couple of million years, approaching the lower limit of the age parameter, meaning the fractional uncertainties of age estimates for these stars will be larger than their older counterparts. As an alternative, we can use initial masses as age approximations. Stellar mass estimates are not spaced by any particular resolution, but are instead sampled from a Kroupa IMF over a range of stellar lifetimes and evolutionary phases, providing a varied range of masses at different ages. We can thereby estimate the average age of a star using the mass-luminosity relation, where the lifespan of a star is inversely correlated with its initial mass, defined as follows: 

$$t_{age} = 10^{10} (M/M_{\odot})^{-2.5}$$ 

For example, an $8 M_{\odot}$ star is expected to live up to 55 Myrs as opposed to a $10 M_{\odot}$ star which will only exist for $\sim$30 Myrs \citep{binney1987}. This relation does not account for mass loss from stellar winds in massive stars, which would only widen the age disparity \citep{Castor1975}. 

\begin{figure}
    \centering
    \includegraphics[width=\linewidth]{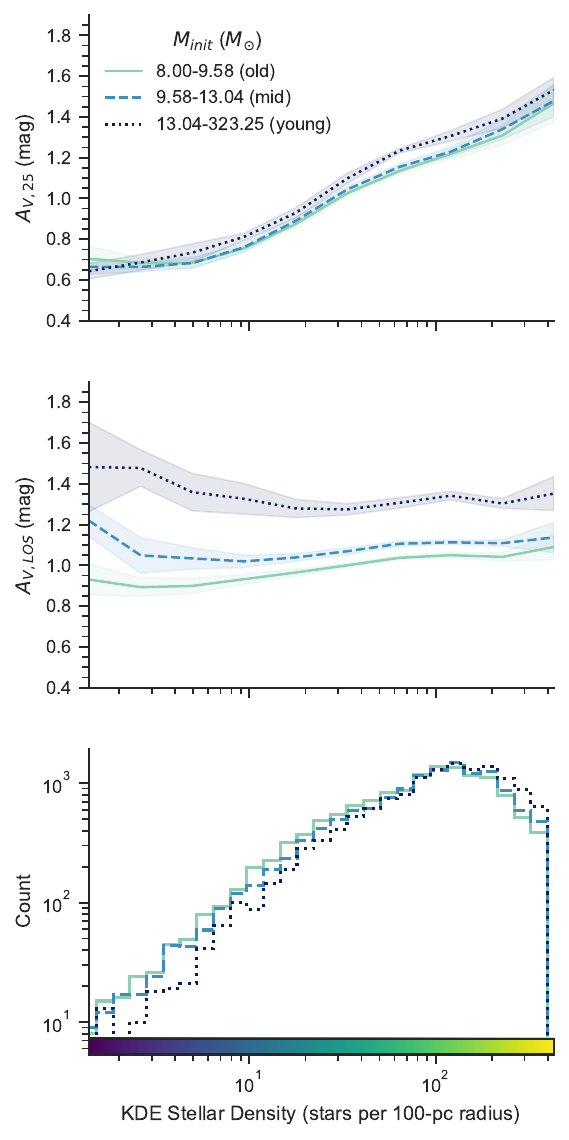}
    \caption{$A_{V,\, 25}$ (top), regional $A_{V,\, LOS}$ (middle), and distribution of massive star candidates as a function of KDE stellar density and initial mass. Mass bins are selected so that they all contain roughly the same number of stars. Uncertainties were calculated using block bootstrapping. Bottom: histogram of mass bin distributions.}
    \label{fig:mass_split}
\end{figure}

In Figure \ref{fig:mass_split}, we split our sources into three equally-sized mass bins and recompute $A_{V,\,25}$ and $A_{V,\,LOS}$ in each initial mass bin. We find no significant difference in the average $A_{V,\, 25}$ (top) as a function of initial mass, indicating that stars sample the same galactic environments regardless of how old they are. However, we find moderate amounts of variation in the $A_{V,\, LOS}$ as a function of initial mass (middle). We discuss the possible interpretations of these variations below. 

Firstly, we find that the average $A_{V,\, LOS}$ increases as a function of initial mass bin i.e. $A_{V,\, LOS}$ decreases with age. These are the results we would expect if pre-supernova feedback mechanisms from the massive stars were clearing the natal ISM in the local vicinity.

We consider the possibility that observational biases might also be contributing to this trend. As we discussed in Section \ref{sec:av_limits}, an extinguished $30 M_{\odot}$ star is more easily observed than an $8 M_{\odot}$ star, especially when embedded in molecular clouds. Since we use initial mass estimates as age proxies, it is possible that sources in the lowest mass bin ($M_{init} = 8.0$--$9.58$ $M_{\odot}$) preferentially less observed at high $A_V$, even if they do exist.  

To test whether the distribution of extinction is dependent on mass, we use the extinguished stellar SED models described in Section \ref{sec:av_limits} to generate a catalog of simulated massive stars with realistic extinctions. To simulate a realistic distribution of extinction, we construct a log-normal distribution using the average $\mu$ and $\sigma$ of total column extinction from our massive star candidate sources ($\mu = 1.3$, $\sigma=0.3$). To simulate a Kroupa IMF, we assume the probability of integer initial masses to be proportional to $M^{-2.35}$. From these two distributions, we random-sample 41,000 massive stars and compare their extinguished magnitudes to the 50\% completeness limits in each band, assuming crowding is not an issue. If a star is detectable in all bands, we store the applied extinction in an array corresponding to the appropriate mass bin, $M_{init} = [8$--$9$, $10$--$13$, $+14$] $M_{\odot}$. From each observed extinction array, we compute the mean $A_V$ for each mass bin: [$1.293\pm 0.325, 1.293\pm 0.333, 1.301\pm 0.346$] mag, respectively. Although there is a slight increase in $A_V$ as a function of mass bin, this change is not enough to account for the large discrepancy we observe in Figure \ref{fig:mass_split} (middle), which differs by 0.25 mag in the densest regions, to nearly 0.6 mag in the sparsest regions. As such, our simulation indicates that the results in Figure \ref{fig:mass_split} do not suffer from observational biases due to extinction. 

We note though that if the tail of dust distribution were larger, (e.g. $\sigma=0.5$), then the observational bias becomes more pronounced, with mean $A_V = [1.258\pm 0.418, 1.298 \pm 0.463$, $1.369\pm0.555$] mag, respectively. This could be the case in denser molecular clouds, however, this bias is still not enough to account for the results in Figure \ref{fig:mass_split}.

Based on this investigation, we find evidence that older massive stars are less embedded in their natal ISM than their younger counterparts. Since we do not see significant changes in the regional extinction as a function of stellar age, we believe the change in $A_{V,\, LOS}$ is due to feedback mechanisms clearing out the natal ISM surrounding the stars on scales smaller than 25-pc. However, based on these results alone, it is unclear whether this decrease in embeddedness is only the product of pre-supernova feedback mechanisms or if supernova feedback from other nearby massive stars contributed as well. One way to differentiate between the two scenarios would be to cross-correlate the spatial distribution of massive stars with the locations of known supernova remnants in M31. General proximity, along with ages and sizes of supernova remnants would determine the likelihood that a massive star is emerging from its natal ISM due to collective feedback effects. Inversely, it is then possible to constrain the effects of isolated pre-supernovae feedback from observations.

\subsection{How does dispersion affect the spatial distribution of massive stars?}

As stars age, they naturally decouple from the ISM due to dispersion. This is evident from the fact that older stellar populations, such as Red Giant Branch (RGB) stars, are more homogeneously distributed across galaxies, whereas younger stars generally trace molecular gas \citepalias[]{dalcanton2015}. We note that decoupling between stars and gas due to dispersion is distinct from decoupling due to feedback, as dispersion occurs at much longer length and time-scales --- kpc compared to pc, Gyrs compared to Myrs. %The spatial decoupling between gas and stars, i.e. the rate of dispersion, can be probed by fitting analytical models to nearby galaxy data \citep{kim2021}. Otherwise, 
Directly determining the rate of dispersion is challenging because you need to know in which cloud stars formed, requiring resolutions only obtainable in the Milky Way. Recent studies of nearby molecular regions find the differential velocity, i.e. the speed at which stars disperse from the ISM, to be on the order of a couple km $s^{-1}$ \citep{swiggum2021, galli2019}. In this section, we explore possible evidence of massive stars in the catalog dispersing over time.

Since dispersion effects compound over longer periods of time, as opposed to feedback, we expect older massive stars (i.e. lower-mass) to be less clustered compared to their younger counterparts (i.e. higher-mass). Using the results from Section \ref{sec:feedback}, we plot the distribution of massive stars as a function of mass bins at the bottom of Figure \ref{fig:mass_split}. We find that the most massive stars tend to exist in denser stellar density environments compared to their less massive counterparts. This is generally what we would expect to find from the effects of dispersion: stars gradually migrate away from each other and the clouds they were born from, moving into more diffuse regions. These results confirm that the massive star candidates in the catalog are dispersing over time.

\subsection{Does extinction vary by galactic radius?}\label{sec:disc_rad}

\begin{figure*}
    \centering
    \includegraphics[width=\linewidth]{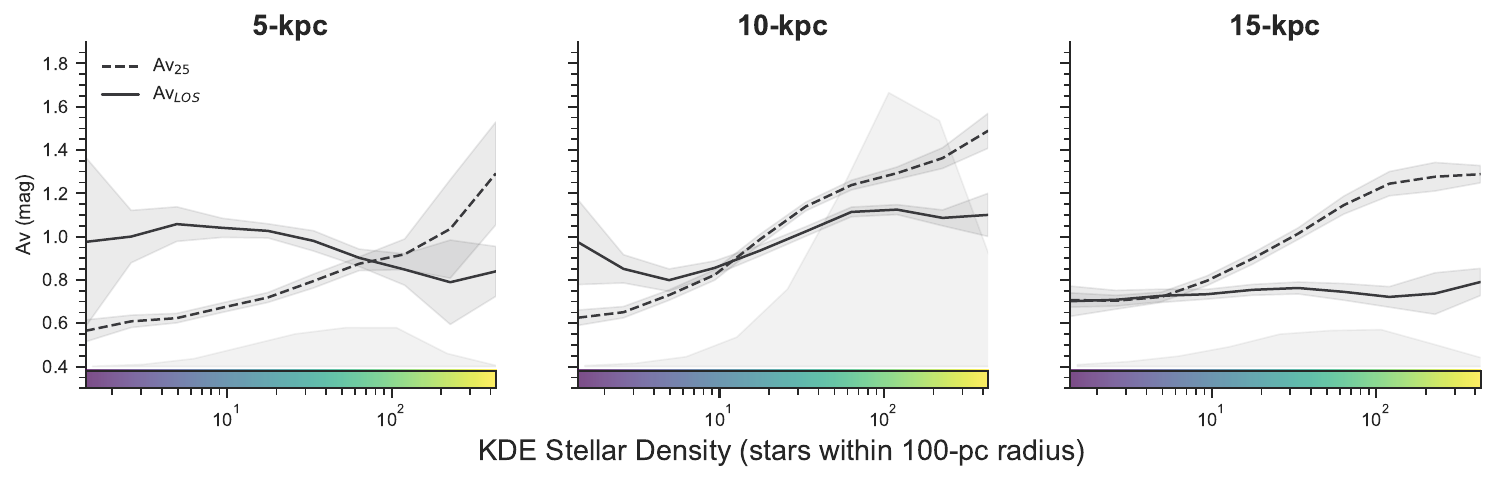}
    \caption{Same as figure \ref{fig:main_cumulative}, but calculated for each individual ring as noted in Figure \ref{fig:kde}.}
    \label{fig:main_rings}
\end{figure*}
 
Massive star formation efficiency has been observed to vary in extreme galactic environments \citep[e.g. the central molecular zone,][]{henshaw2023}, and from recent interactions with other galaxies, however, it is still unclear whether star formation efficiency is influenced by galactic radius \citep{kennicutt2012, lin2020}. Although M31 has interacted with satellites in the past (e.g. M32, M33), the most recent interactions are thought to have transpired over 2 Gyrs ago \citep{dsouza2018}. As such, we do not expect M31's current star formation rate to be enhanced due to tidal interaction, with observations showing low recent star formation rate intensities \citep{williams2017}. 

M31 has three major spiral arms, which are actually suspected to be rings, $\sim$5, 10, and 15 kpc away from the center \citep[][]{gordon2006}. To distinguish between galactic environment and radial distance, we split the catalog into three separate radial rings and examine how massive star extinction properties change as a function of galactic radius. We separate each ring by splitting sources at 6.7 kpc and 13.0 kpc (Figure \ref{fig:kde}). We find 5,153, 31,196, and 5,758 sources in the 5, 10, and 15-kpc rings, respectively. Figure \ref{fig:main_rings} shows the average $A_{V,\, LOS}$ and $A_{V,\, 25}$,  now recomputed as a function of radial distance.

We find that the 10 and 15-kpc rings display similar trends as Figure \ref{fig:main_cumulative}, however, the 5-kpc ring has significantly less regional extinction, despite being more centrally located within the galaxy. Similar results were noted by \citet{lewis2015} who found that the 5-kpc ring has not experienced much star formation in the last 500 Myrs, longer than the average lifetime of a massive star. Our results agree with \citet{lewis2015}, as we find fewer massive stars within the 5-kpc ring compared to the 10 and 15-kpc rings, although the reasoning for this deficiency is still not well understood. %This is in stark contrast to the overall stellar surface density of RGB sources which \citetalias{dalcanton2015} found to decrease radially away from the center.

\section{Future Work}\label{sec:followup}

One potential option for validating in situ formation of massive stars in low-density environments would be to observe gas tracers at select locations with the James Webb Space Telescope (JWST). Photometry with NIRCam or MIRI would provide high-resolution imaging of dust (sub-1 pc resolution at the distance of M31), allowing us to verify whether smaller dust structures exist near massive stars in regions with low stellar density. Potential tracers include polycyclic aromatic hydrocarbons (PAHs), small carbonaceous dust grains that make up a small percentage of the total dust mass. When these grains absorb UV photons, they heat up and emit at various wavelengths in the infrared, more notably at 3.3 $\mu$m, 7.7 $\mu$m, and 11.3 $\mu$m, which are observable with the NIRCam and MIRI instruments. PAH luminosity is proportional to the radiation field intensity around them, meaning their emission can be used to trace shielding from molecular clouds. Alternatively, $H_2$ emission is observable via the 1-0 S(1) ro-vibrational line at 2.12 $\mu$m, a small fraction of which is excited when exposed to high radiation fields. As such, 2.12 $\mu$m emission can be used to trace the surface density of molecular structures at sub-pc scales. While studies of $H_2$ with JWST have so far mainly been located in the Milky Way \citep{berne2022}, PAHs have been used in extragalactic studies to show star formation occurring in smaller spurs located between spiral arms \citep[e.g.,][]{williams2022}.

\section{Conclusion}\label{sec:conc}

% The results paint an interesting picture of the state of the ISM in the immediate vicinity of massive stars. 

% %

% %Past a certain limit, a young embedded star will become so red and dim that it shifts beyond our observational limits with HST, both in magnitude and color, and is only detectable in IR. While this jump in line-of-sight extinction could have been justified by the explanation that we are merely observing more embedded stars, the sheer volume of highly embedded stars would be completely indefensible. This is more likely due to the fact that AGB stars were misclassified as extremely massive stars with exceptionally high line-of-sight extinction. 

% This is evident from the fact that massive stars across all stellar densities have similar line-of-sight extinction ($A_{V,\, LOS}$), despite the fact that stars in less dense stellar environments have very low amounts of total column density extinction at 25-pc resolution ($A_{V,\, 25}$). This indicates that regions with low amounts of total column density extinction are still able to reach a volume density sufficient for star formation and that not all star-forming molecular clouds are resolvable at 25-pc.

In this paper, we investigated the immediate environments of massive stars compared to the broader ISM in which they are situated. We find that the line-of-sight extinction towards massive stars stays constant on average, regardless of location in the galaxy. These results suggest that massive stars are forming in-situ in low-density environments. %However, high-resolution follow-up observations of interstellar gas or dust tracers are required to confirm this.

Our conclusions are as follows:

\begin{itemize}
    \item We construct a catalog of 42,107 massive star candidates in the PHAT region of M31 using multi-band photometric observations fit with the Bayesian Extinction and Stellar Tool (BEAST). We compare our candidate catalog with spectroscopic OB stars from \citet{massey2016} and find acceptable overlap between the observed magnitudes. In addition, the spatial distribution of the candidates tightly traces 24-micron emission from dust heated by energetic stars.
    \item We define a metric to quantify the stellar density of massive stars using Kernel Density Estimation with a bandwidth of 100 pc i.e. massive stars per 100-pc radius. With this metric, we find that the environment of massive stars range from 1-400 massive stars per 100-pc radius, with most massive stars observed in environments with an average of $80-200$ massive stars per 100-pc radius. 
    \item We observe the median total column density of dust ($A_{V,\, 25}$, calculated at 25-pc resolution), increases monotonically and logarithmically as a function of stellar density, ranging from an average of $0.7$ mag in the sparsest stellar regions, to $1.6$ mag in the densest stellar regions in the galaxy (Figure \ref{fig:main_cumulative}).
    \item We observe the average line-of-sight extinction ($A_{V,\, LOS}$, derived from SEDs with an equivalent resolution of $\sim$0.5 pc), remains relatively constant as a function of stellar density, with an average $A_V = 1.1$ mag (Figure \ref{fig:main_cumulative}) 
    \item We compute the average flux at the location of massive stars for atomic and molecular gas tracers as a function of stellar density. We find that predicted CO (J=3-2) line flux observed from the HASHTAG survey (resolution of $\sim$30 pc) increases exponentially as a function of stellar density, with CO being completely absent at sparse regions (Figure \ref{fig:co32_coverage}). We find that HI integrated intensity from the VLA (resolution of $\sim$32 pc) also increases as a function of stellar density similar to the total column density extinction (Figure \ref{fig:HI_coverage}). 
    % \item Our findings suggest that massive stars are forming in situ in low-density environments, often regarded as not conducive to star formation. 
\end{itemize}

\begin{acknowledgments}
The authors would like to thank Philip Massey and Eric Koch for their helpful contributions to ancillary data sets that enriched this investigation. The authors acknowledge Interstellar Institute's program ``With Two Eyes`` and the Paris-Saclay University's Institut Pascal for hosting discussions that nourished the development of the ideas behind this work. This research has made use of NASA's Astrophysics Data System Bibliographic Services. The Flatiron Institute is supported by the Simons Foundation.
\end{acknowledgments}

\bibliography{sample631}{}
\bibliographystyle{aasjournal}


\appendix

\section{Extra KDE plots}\label{app:kde_apdx}

We recalculate our results using KDE stellar densities computed using 30 and 300 pc bandwidths. The main difference between these results is that the distribution of KDE stellar densities are severely skewed to sample either low stellar densities (30-pc) or high stellar densities (300-pc) due to the inherent nature of how the metric is constructed.

% \begin{figure*}[b]
%     \centering
%     \includegraphics[width=0.8\textwidth]{./kde_all.pdf}
%     \caption{Left: KDE map of stars using varying bandwidths of 30, 100, and 300-pc. Right: Zoom-in on subregion in the rings.}
%     \label{fig:kde_apdx_map}
% \end{figure*}

\begin{figure*}
    \centering
    \includegraphics[width=\textwidth]{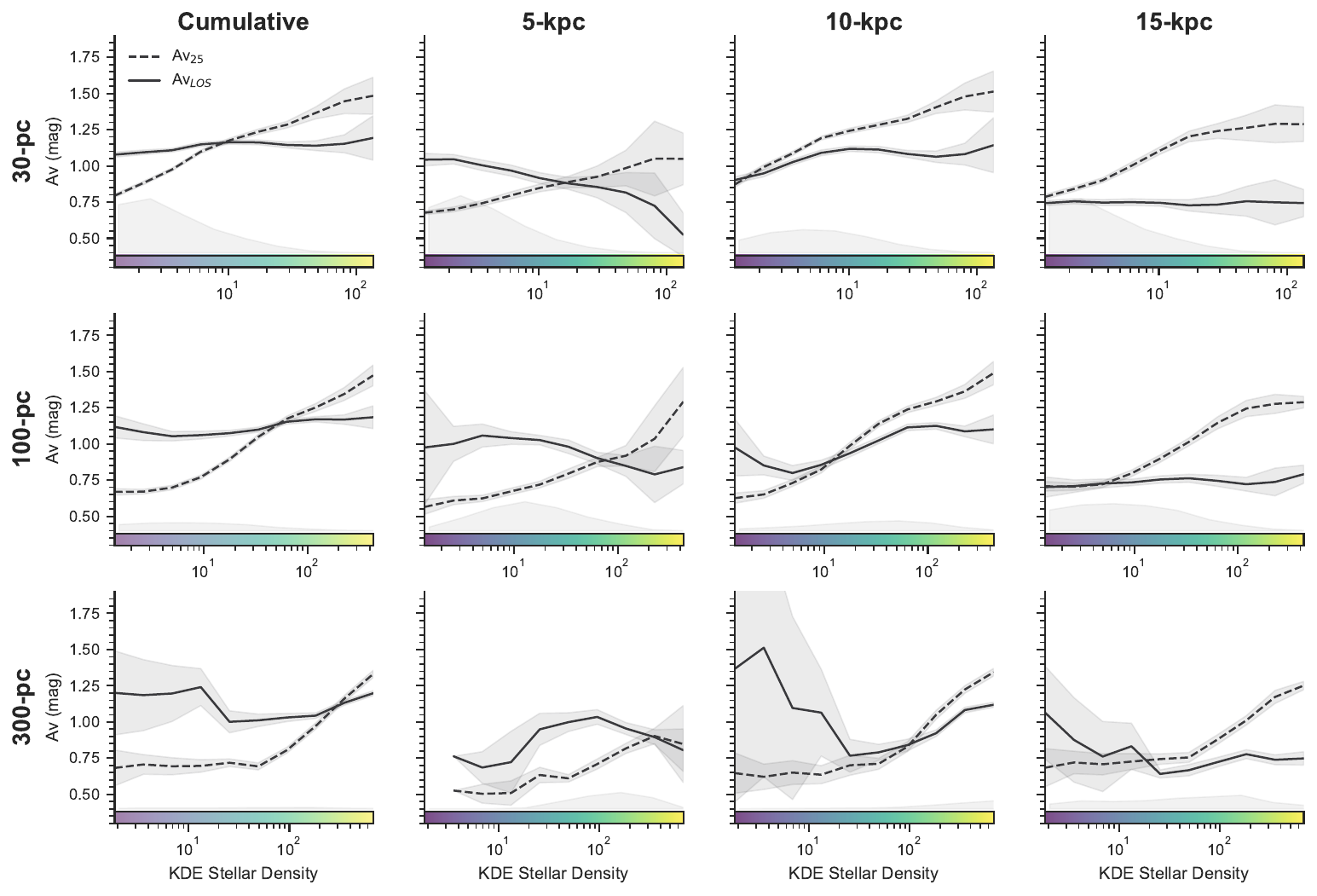}
    \caption{Left: Average regional 25-pc (dashed) and line-of-sight extinction (solid) binned by KDE stellar density at 30, 100, and 300-pc bandwidths. We use block bootstrapping (100x100 blocks, reps=100) to calculate the standard error (shaded). A normalized histogram of the stellar density distribution is plotted along the x-axis. Middle and Right: Cumulative plots split into individual rings.}
    \label{fig:kde_apdx_results}
\end{figure*}



\end{document}